\documentclass[preprint, 1p]{elsarticle}
\usepackage[utf8]{inputenc}
\usepackage{fullpage}
\usepackage{setspace}
\usepackage{amsmath,amssymb,amsfonts}
\usepackage{algorithmic}
\usepackage{graphicx}
\usepackage{algorithm,algorithmic}
\usepackage{hyperref}

\usepackage{textcomp}
\usepackage{latexsym,amscd, bbm}
\usepackage[mathscr]{euscript}
\usepackage{bigints}
\DeclareMathOperator*{\argmin}{arg\,min}

\usepackage{float}
\usepackage{caption}
\usepackage{breqn}
\usepackage{color}
\usepackage{arydshln}
\usepackage{bm}
\graphicspath{{./Figures/}}

\newtheorem{thm}{Theorem}

\newtheorem{prop}[thm]{Proposition}
\newdefinition{rmk}{Remark}
\newproof{pf}{Proof}
\newproof{pot}{Proof of Theorem \ref{thm2}}
\usepackage{algorithm}

\begin{document}
\title{Co-Clustering Multi-View Data Using the Latent Block Model}

\author[1]{Joshua Tobin\corref{cor1}}
\ead{tobinjo@tcd.ie}
\author[2]{Michaela Black}
\author[1]{James Ng}
\author[2]{Debbie Rankin}
\author[7]{Jonathan Wallace}
\author[3]{Catherine Hughes}
\author[3]{Leane Hoey}
\author[4]{Adrian Moore}
\author[2]{Jinling Wang}
\author[3]{Geraldine Horigan}
\author[5]{Paul Carlin}
\author[3]{Helene McNulty}
\author[6]{Anne M Molloy}
\author[1]{Mimi Zhang}

\affiliation[1]{organization={School of Computer Science \& Statistics, Trinity College Dublin},
city={Dublin},
country={Ireland}}
\affiliation[2]{organization={School of Computing, Engineering \& Intelligent Systems, Ulster University},
city={Derry $\sim$ Londonderry},
country={United Kingdom}}
\affiliation[7]{organization={School of Computing, Ulster University},
city={Belfast},
country={United Kingdom}}
\affiliation[3]{organization={School of Biomedical Sciences, Nutrition Innovation Centre for Food and Health, Ulster University},
city={Coleraine},
country={United Kingdom}}
\affiliation[4]{organization={School of Geographic \& Environmental Sciences, Ulster University},
city={Coleraine},
country={United Kingdom}}
\affiliation[5]{organization={School of Health, Wellbeing \& Social Care, The Open University},
city={Belfast},
country={United Kingdom}}
\affiliation[6]{organization={School of Medicine, Trinity College Dublin},
city={Dublin},
country={Ireland}}

\begin{abstract}
The Latent Block Model (LBM) is a prominent model-based co-clustering method, returning parametric representations of each block cluster and allowing the use of well-grounded model selection methods. The LBM, while adapted in literature to handle different feature types, cannot be applied to datasets consisting of multiple disjoint sets of features, termed views, for a common set of observations. In this work, we introduce the multi-view LBM, extending the LBM method to multi-view data, where each view marginally follows an LBM. In the case of two views, the dependence between them is captured by a cluster membership matrix, and we aim to learn the structure of this matrix. We develop a likelihood-based approach in which parameter estimation uses a stochastic EM algorithm integrating a Gibbs sampler, and an ICL criterion is derived to determine the number of row and column clusters in each view. To motivate the application of multi-view methods, we extend recent work developing hypothesis tests for the null hypothesis that clusters of observations in each view are independent of each other. The testing procedure is integrated into the model estimation strategy. Furthermore, we introduce a penalty scheme to generate sparse row clusterings. We verify the performance of the developed algorithm using synthetic datasets, and provide guidance for optimal parameter selection. Finally, the multi-view co-clustering method is applied to a complex genomics dataset, and is shown to provide new insights for high-dimension multi-view problems. 

\end{abstract}


\maketitle
\section{Introduction}
Clustering algorithms help to provide a global overview of a dataset. In cases where the number of features is large, just as it is necessary to summarize the individuals into homogeneous groups, it is useful to summarize the features. This simultaneous clustering of instances and features is referred to as \textit{co-clustering}. A large and complex data matrix can thus be summarized by a limited number of blocks, corresponding to the intersection of the row and column clusters. 

The Latent Block Model (LBM) \citep{govaert_clustering_2003, govaert_em_2005} is a model-based co-clustering method that models the elements of a block cluster using a parametric distribution. This makes the block interpretable through the distribution's parameters. Additionally, model selection methods like the Integrated Completed Likelihood (ICL) can be employed to determine the suitable number of row and column clusters. Despite demonstrating effectiveness and being extended to handle continuous, ordinal, and categorical data independently and simultaneously in \citep{selosse_model-based_2020}, the LBM method is constrained in its analysis, providing only a single common grouping of sample participants across all data matrices. This limitation becomes problematic as the experimenters now regularly collect data from multiple modalities to allow phenomena of interest to be investigated from several perspectives. Such an integrative study requires a clustering method that can take as input multi-view data, namely a fixed set of observations with several disjoint sets of features. 

We here develop an approach, Multi-View Latent Block Model (MVLBM), that extends the latent block model to the multi-view context. Following the multi-view mixture model method of \citet{carmichael_learning_2020}, our model operates under two key assumptions for datasets with $V \geq 2$ views:
\begin{enumerate}
    \item Marginally, each view follows a latent block model, i.e. there are $V$ sets of of view-specific row and column clusters. 
    \item The views are independent when conditioned on the row cluster memberships of the respective marginal views.
\end{enumerate}

\begin{figure}[!ht]
    \centering
    \includegraphics[width=0.9\linewidth]{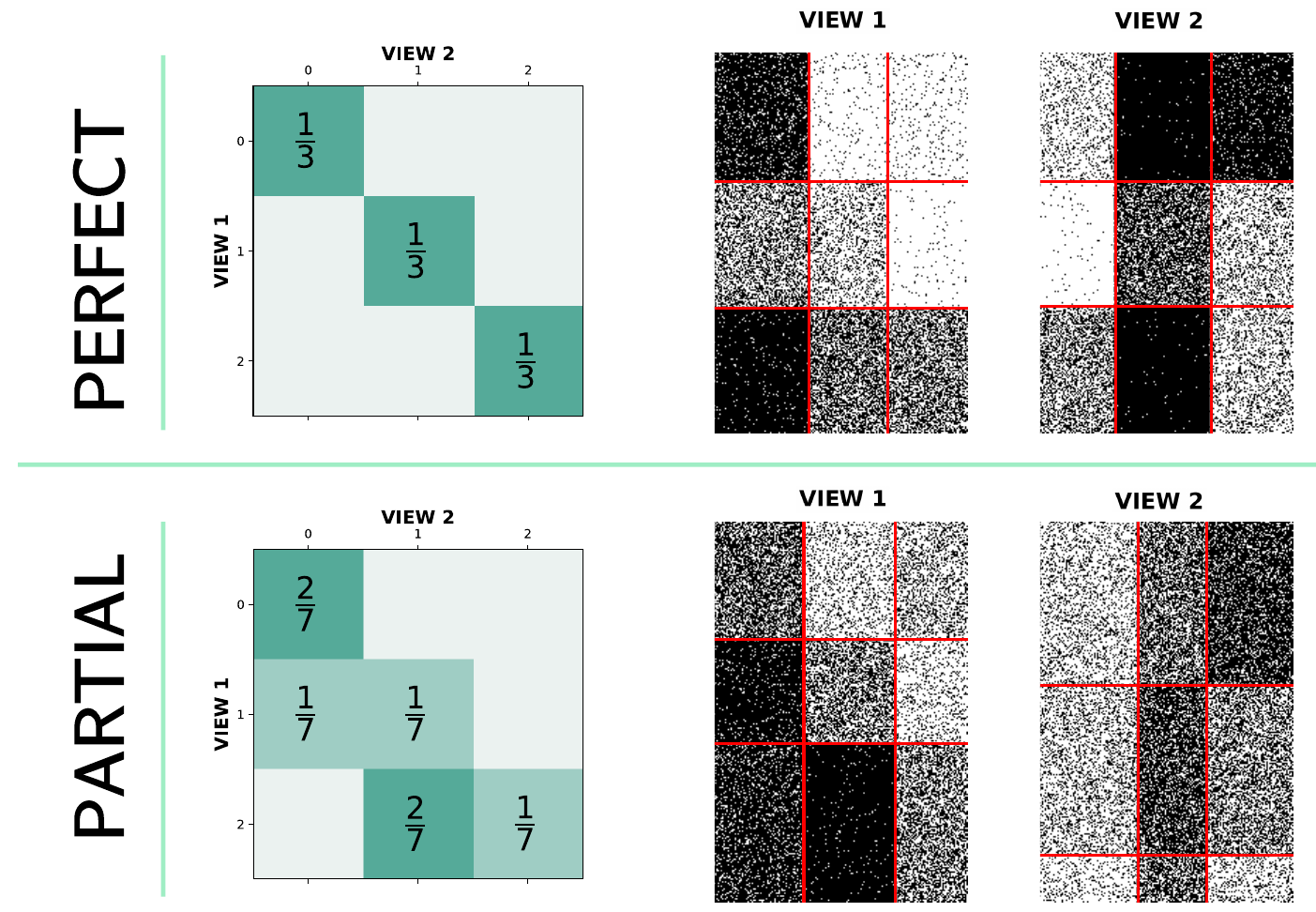}
    \caption{The row cluster probability matrix $\bm{\pi}$ captures the dependency between the row clusters in each view. For a $V=2$ view dataset, a comparison between perfect (top) and partial (bottom) dependency structures is presented. In both cases, there is a relationship between the row clusters in each view. For the perfect case, the partitions are identical. For the partial case, membership of a particular cluster in one view is predictive of the assignment in the second view. }
    \label{fig:dependence}
\end{figure}

It is assumed that there is potentially a partial dependence between the row clusters in different views. In a two-view dataset, every instance has two cluster label vectors $\bm{z}_1 \in \mathbb{R}^{n \times K_1}$ and $\bm{z}_2 \in \mathbb{R}^{n \times K_2}$, where $K_{v}$ is the number of row clusters in the $v$th view; $z_{ik_1} = 1$ if $\bm{x}_i$ is in row cluster $k_1$ in the first view and is 0 otherwise, and $z_{ik_2} = 1$ if $\bm{x}_i$ is in row cluster $k_2$ in the second view and 0 otherwise. We are interested in the dependency between the row cluster assignments across the two views. In place of a cluster membership probability vector, capturing the prior probability of an instance belonging to a row cluster in a data view, in the multi-view case the joint distribution of the cluster labels is described by the row cluster probability matrix $\bm{\pi} \in \mathbb{R}_{+}^{K_{1} \times K_{2}}$. The structure of the matrix $\bm{\pi}$ captures the dependency between the row cluster assignment across the two views:
\begin{equation*}
    \pi_{k_1k_2} = \frac{1}{n} \sum_{i = 1}^n \Pr(z_{ik_1} = 1, z_{ik_2} = 1).
\end{equation*}
Thus the joint probability of the row cluster labels in the view is computed as: 
\begin{equation*}
    \Pr(\bm{z}_1, \bm{z}_2) = \prod_{i, k_1, k_2} \pi_{k_1k_2}^{z_{ik_1} z_{ik_2}}. 
\end{equation*}
An example of different dependency structures for a $V = 2$ view dataset is provided in Figure \ref{fig:dependence}. As the sets of features in the different views are disjoint and distinct, we assume no relationship between them across views. Therefore, our objective is to concurrently learn the structure of $\bm{\pi}$ and the block-cluster parameters for each view using a latent block model.

Jointly estimating row cluster memberships across views naturally prompts consideration of the association between the underlying clusterings in each view. To assess this association, we introduce a hypothesis testing procedure adapted from \cite{gao_are_2020, gao2022testing}. This test evaluates the null hypothesis that clusterings on two views of a single set of observations are independent. The results of this test are then used to inform the estimation procedure of the MVLBM. 

Increasing the number of views, and the number of row clusters in each view, results in exponential growth of the joint cluster space. In extreme cases, this can lead to empty clusters, posing challenges for likelihood-based estimation methods. To encourage sparsity in the row clusters, we introduce a penalized likelihood approach, adapting a logarithmic penalty from \cite{huang_model_2017, carmichael_learning_2020}. The approach acts as a threshold, removing clusters with mixing proportions below a user-specified value.

The paper is organized as follows: Section \ref{sec:s2} reviews relevant literature on LBMs and multi-view co-clustering. Section \ref{sec:s3} provides a detailed description of the MVLBM method. The motivation and explanation of the hypothesis testing procedure are presented in Section \ref{sec:s4}. Section \ref{sec:s5} introduces a penalized likelihood approach aimed at promoting sparsity in row clusterings. The efficacy of the MVLBM approach is validated through extensive simulated analyses in Section \ref{sec:s6}. In Section \ref{sec:s7}, MVLBM is applied to a challenging problem in multi-view genomics analysis. Finally, Section \ref{sec:s8} concludes with a summary of findings and an outline of potential future directions.

\section{Related Work}\label{sec:s2}
\subsection{Latent Block Model}\label{sec:s2s1}
The LBM of \citep{govaert_clustering_2003, govaert_block_2008} is the leading model-based co-clustering algorithm. The LBM extends the classical mixture-model framework of \citep{banfieldModelBasedGaussianNonGaussian1993b} by considering the assignments of both the rows and columns of the dataset into respective clusters. It assumes that the row and column partitions are independent and that, conditionally on the row- and column cluster membership, the observed random features are also independent realizations of a probability distribution defining the block-cluster. While originally developed solely for Bernoulli mixture models, the approach has been extended to encompass Gaussian data \citep{nadifAlgorithmsModelbasedBlock2008}, ordinal data \citep{jacques_model-based_2018, selosse_analysing_2019}, functional data \citep{bouveyron_co-clustering_2022}, and time-dependent data \citep{casa_co-clustering_2021}. 


A popular extension of the LBM integrates constraints, allowing users to restrict which features are grouped together to form column clusters \cite{robert2017classification}. In \citep{selosse_analysing_2019} and \citep{selosse_model-based_2020}, this restricted model is applied allowing the dataset to contain features that do not share a common support, e.g. ordinal features with different numbers of levels, or both continuous and categorical features. The constrained LBM approach could be considered for the multi-view problem, treating the features from each view as a set with constraints, preventing any features from two views to be clustered together. However, such an approach is equivalent to consensus clustering, enforcing one clustering of the rows to be shared across all the views in the dataset. 

\subsection{Multi-View Co-Clustering}\label{sec:s2s2}
While no LBM method has, to date, been developed specifically for the multi-view co-clustering problem, two related fields of research yield insights to the problem. 

Firstly, several multi-view co-clustering methods have been developed that extend and apply discriminative co-clustering methods to the multi-view problem. \citet{tokuda2017multiple} introduce a non-parametric Bayesian mixture model approach, which is applicable to data containing heterogeneous feature types. \citet{sunMultiviewSparseCoclustering2015} repeatedly approximate the data matrices in each view using sparse rank 1 approximations, thus decomposing the matrix into a pair of left and right vectors. The non-zero components of each of the vectors correspond to memberships of the row and column clusters, respectively. The authors enforce a consensus clustering of the rows across the views. Another matrix factorization approach is offered in \citet{nieAutoweightedMultiviewCoclustering2020}. Spectral co-clustering methods are applied in \citet{huangAutoweightedMultiviewCoclustering2020} with bipartite graphs computed by stacking the Laplacian representations from each view. Such approaches provide only a hard clustering of the data, and further fail to provide useful representatives of each block cluster in the form of distributional parameters, as done by the LBM. 

Secondly, mixture model clustering of row observations has been extended to the multi-view setting. The method of \citet{bickelMultiviewClustering2004} introduces mixture modelling for multiple data views, while enforcing the requirement that there is one consensus clustering of the instances that is present across all views. This work ignores how clustering information is partially shared by latent signals across the views. The works of \citet{gao_are_2020} and \cite{gao2022testing} make important contributions, developing tests for independence between the clusters in a two-view mixture model and two-view network data, respectively. The multi-view mixture model on which this work is based is developed in \citet{carmichael_learning_2020}. There, two novel methods are introduced exploring how information is shared between views. The methods impose interpretable structures on the cluster membership matrix, firstly using sparsity and secondly through the enforcement of block-diagonal constraints.

\section{Multi-view Latent Block Models}\label{sec:s3}
\subsection{Latent Block Model}\label{sec:s3s1}
We begin with a description of the single-view LBM as introduced in \citep{govaert_block_2008}. Consider the data matrix $\bm{x} = (x_{ij})_{ij}$ where $i \in \lbrace 1, \ldots, n \rbrace$ and $j \in \lbrace 1, \ldots, d \rbrace$. It is assumed that there are $K$ row clusters and $L$ column clusters that correspond to a partition $\bm{z} = (z_{ik})_{ik}$ of the rows and a partition $\bm{w} = (w_{jl})_{jl}$ of the columns, where $k \in \lbrace 1, \ldots, K \rbrace$ and $l \in \lbrace 1, \ldots, L \rbrace$. Here, $z_{ik} = 1$ if row $i$ belongs to row cluster $k$, and 0 otherwise; $w_{jl} = 1$ if column $j$ belongs to column cluster $l$, and 0 otherwise. In order to simplify the notations, the underlying range of variation will be omitted in the sums and products. As such, $\sum_{i=1}^n, \sum_{j=1}^d, \sum_{k=1}^K, \sum_{l=1}^L$ and $\prod_{i=1}^n, \prod_{j=1}^d, \prod_{k=1}^K, \prod_{l=1}^L$ will be written as $\sum_i, \sum_j, \sum_k, \sum_l$ and $\prod_i, \prod_j, \prod_k, \prod_l$, respectively.

The first assumption of the LBM is that the univariate random features $x_{ij}$ are conditionally independent given the row and column partitions $\bm{z}$ and $\bm{w}$. Therefore, the conditional probability density of $\bm{x}$ given $\bm{z}$ and $\bm{w}$ is 
\begin{equation*}
    p(\bm{x}| \bm{z}, \bm{w}; \bm{\alpha}) = \prod_{i,j,k,l} p(x_{ij}; \alpha_{kl})^{z_{ik}w_{jl}},
\end{equation*}
where $\alpha_{kl}$ are the distribution parameters of the block $(k, l)$. Furthermore, we denote the set of parameters for all blocks as $\bm{\alpha} = \lbrace \alpha_{kl}: k \in 1, \ldots, K, ~~ l \in 1, \ldots, L \rbrace$.

Secondly, the LBM assumes that the latent variables $\bm{z}$ and $\bm{w}$ are independent, so $p(\bm{z}, \bm{w}; \bm{\pi}, \bm{\rho}) = p(\bm{z}; \bm{\pi})p(\bm{w}; \bm{\rho})$ with 
\begin{equation*}
    p(\bm{z}; \bm{\pi}) = \prod_{i,k} \pi_{k}^{z_{ik}} ~~~~~ \text{and} ~~~~~p(\bm{w}; \bm{\rho}) = \prod_{j,l} \rho_{l}^{w_{jl}},
    \end{equation*}
where $\pi_{k}  = \Pr (z_{ik}= 1)$ and $\rho_{l} = \Pr (w_{jl} = 1)$. This implies that for all $i$, the distribution of $\bm{z}_i$ is the multinomial distribution $\mathcal{M}(1, \bm{\pi})$ and does not depend on $i$. Similarly, for all $j$, the distribution of $\bm{w}_j$ is the multinomial distribution $\mathcal{M}(1, \bm{\rho})$ and does not depend on $j$.  As a result, the parameters for the LBM are defined as $\bm{\Theta} = (\bm{\pi}, \bm{\rho}, \bm{\alpha})$. Therefore, if $Z$ and $W$ are the sets of all possible labels $\bm{z}$ and $\bm{w}$ respectively, the probability density function of $\bm{x}$ is 
\begin{align*}
p(\bm{x}; \bm{\Theta}) &= \sum_{(\bm{z} \times \bm{w}) \in Z \times W} p(\bm{z}; \bm{\Theta}) p(\bm{w}; \bm{\Theta}) p(\bm{x}|\bm{z}, \bm{w}; \bm{\Theta}) \\
&= \sum_{(\bm{z} \times \bm{w}) \in Z \times W} p(\bm{z}; \bm{\pi}) p(\bm{w}; \bm{\rho}) p(\bm{x}|\bm{z}, \bm{w}; \bm{\alpha}) \\
&= \sum_{(\bm{z} \times \bm{w}) \in Z \times W} \prod_{i,k} \pi_{k}^{z_{ik}} \prod_{j,l} \rho_{l}^{w_{jl}} \prod_{i,j,k,l} p(x_{ij}; \alpha_{kl})^{z_{ik}w_{jl}}.
\end{align*}

A key issue in the application of LBMs is the impracticality of estimation using the EM algorithm. Specifically, the E-step requires the calculation of the joint conditional distribution of the missing labels, which cannot be factorized due to the conditional dependence between the row and column labels and the observations. Several alternative estimation approaches have been proposed in the literature. The variational method of \citep{govaert_block_2008} assumes the joint distribution of the row and column labels can be factored, using the variational approximation, thus allowing row-wise and column-wise EM algorithms to be iterated. The SEM-Gibbs approach of \citep{govaert_co-clustering_2013} reduces the sensitivity to initial values by simulating the latent variables according to their conditional probability using Gibbs sampling. Bayesian approaches, using collapsing conjugate priors, have also been proposed for Bernoulli and Gaussian type features \citep{wyse_block_2012}.

\subsection{Multi-View Latent Block Model}\label{sec:s3s2}
We now describe the multi-view LBM for $V \geq 2$ views. The model assumes that each view follows an LBM, and that the views are conditionally independent given the cluster memberships. To simplify notation in what follows, the view index is provided as a subscript for terms with no subscript index, e.g. $\bm{x}_{v}$ refers to the data matrix for the $v$th view, whereas no additional view index is provided when the view index is represented in existing subscript indices, i.e., $x_{ij_{v}}$ refers to the object in the $i$th row and $j_{v}$th column of $\bm{x}_{v}$ and $\bm{\alpha}_{k_{v}l_{v}}$ refers to the parameters of the $(k_{v}, l_{v})$ block in the $v$th view clustering. 

Let $\bm{x}_{v} = (x_{ij_{v}})_{ij_{v}} \in \mathbb{R}^{n \times d_{v}}$ denote the data matrix in the $v$th view, and $\bm{x}_{v_{i\cdot}}$ denote the $i$th row of $\bm{x}_v$. Each view has $n$ observations, while the number of features $d_{v}$ may be different in each view. We assume that there are $K_{v}$ row clusters and $L_{v}$ column clusters in the $v$th view. We denote by $\bm{z}_{v} = (z_{ik_{v}})_{ik_{v}}$ the view-specific row cluster memberships for the $v$th view. In the multi-view setting, we are also interested in the joint distribution of the row cluster labels across all of the views. As before, we denote the row cluster membership multi-array as $\bm{\pi} \in \bm{\Delta}^{K_{1} \times \cdots \times K_{V}}$, with 
\begin{equation*}
    \pi_{k_{1} \ldots k_{V}} = \Pr(z_{ik_1} = 1, \ldots, z_{ik_V} = 1).
\end{equation*}

It is assumed that the row cluster memberships follow a joint distribution 
\begin{equation}\label{eq:s3eq1}
 p(\bm{z}_{1}, \ldots, \bm{z}_{V} ; \bm{\pi}) =\prod_{i,k_1,\ldots, k_V} \pi_{k_{1} \ldots k_{V}}^{z_{ik_1}\times \cdots \times z_{ik_V}}, 
\end{equation}
where $\bm{\pi} \in \bm{\Delta}^{K_{1} \times \cdots \times K_{V}}$ is the row cluster membership multi-array, with $\pi_{k_{1} \ldots k_{V}} = \Pr(z_{ik_1} = 1, \ldots, z_{ik_V} = 1)$. Let $\bm{w}_{v}$ denote the view specific column cluster memberships for the $v$th view. The column clusters are considered disjointly across each view. As such, for the $v$th view
\begin{equation}\label{eq:s3eq2}
    p(\bm{w}_{v}; \bm{\rho}_v) = \prod_{j_{v},l_{v}} \rho_{l_{v}}^{w_{j_{v}l_{v}}},
\end{equation}
where $ \bm{\rho}_{l_{v}} = \Pr(w_{j_{v}l_{v}} = 1)$. As before, we assume that the latent variables for the row and column clusters are independent, so $p(\bm{z}_1, \ldots, \bm{z}_V, \bm{w}_v; \bm{\pi}, \bm{\rho}_v) = p(\bm{z}_1, \ldots, \bm{z}_V; \bm{\pi})p(\bm{w}_v;\bm{\rho}_v)$. Within a specific view, the conditional probability of $\bm{x}_{v}$ given $\bm{z}_{v}$ and $\bm{w}_{v}$ is 
\begin{equation}\label{eq:s3eq3}
    p(\bm{x}_{v}| \bm{z}_{v}, \bm{w}_{v}; \bm{\alpha}_{v}) = \prod_{i, j_v, k_v, l_v} p(x_{ij_{v}}; \alpha_{k_{v}l_{v}})^{z_{ik_{v}}w_{jl_{v}}} ~~~ \\\text{ for } k_{v} \in \lbrace1, \ldots, K_{v} \rbrace, l_{v} \in  \lbrace 1, \ldots, L_{v} \rbrace, v \in \lbrace 1, \ldots V \rbrace.
\end{equation}



The parameters for the MVLBM are defined as $\bm{\Theta} = (\bm{\pi}, \bm{\theta}_{1}, \ldots,\bm{\theta}_{V})$, where $\bm{\pi}$ captures the dependency between the views and $\bm{\theta}_v = (\bm{\rho}_v, \bm{\alpha}_v)$ captures the within-view parameters. Taking $Z_v$ and $W_v$ to be the sets of all possible labels for $\bm{z}_v$ and $\bm{w}_v$ respectively, the probability density function of $\bm{x}$ is 
\begin{equation}\label{eq:s3eq4}
    p(\bm{x}; \bm{\Theta}) = \sum_{\left( (\bm{z}_v) \times (\bm{w}_v) \right) \in \left(\bm{Z}_v \times \bm{W}_v\right)  }p(\bm{z}_{1}, \ldots, \bm{z}_{V}; \bm{\Theta})\prod_{v} p(\bm{w}_{v}; \bm{\Theta}) p(\bm{x}| \bm{z}_{1}, \ldots, \bm{z}_{V}, \bm{w}_{v}; \bm{\Theta}),
\end{equation}
or, equivalently,
\begin{equation*}
    p(\bm{x}; \bm{\Theta}) = \sum_{\left( (\bm{z}_v) \times (\bm{w}_v) \right) \in \left(\bm{Z}_v \times \bm{W}_v\right)  }p(\bm{z}_{1}, \ldots, \bm{z}_{V}; \bm{\pi})\prod_{v} p(\bm{w}_{v}; \bm{\rho}_v) p(\bm{x}| \bm{z}_{1}, \ldots, \bm{z}_{V}, \bm{w}_{v}; \bm{\alpha}_v),
\end{equation*}
where 
\begin{align*}
    p(\bm{z}_{1}, \ldots, \bm{z}_{V}; \bm{\Theta}) & =  p(\bm{z}_{1}, \ldots, \bm{z}_{V}; \bm{\pi})\\
     & = \prod_{i,k_1, \ldots, k_V} \pi_{k_{1} \ldots k_{V}}^{z_{ik_1}\times \cdots \times z_{ik_V}}, \\
    p(\bm{w}_{v}; \bm{\Theta}) & = p(\bm{w}_{v}; \bm{\rho}_v) \\
    & = \prod_{j_{v},l_{v}} \rho_{l_{v}}^{{w_{j_{v},l_{v}}}},\\
    p(\bm{x}| \bm{z}_{1}, \ldots, \bm{z}_{V}, \bm{w}_{v}; \bm{\Theta}) & = p(\bm{x}| \bm{z}_{1}, \ldots, \bm{z}_{V}, \bm{w}_{v}; \bm{\alpha}_v)\\ 
    &=\prod_{i,k_{1}, \ldots, k_{V},j_{v},l_{v}} p (x_{ij_{v}}; \alpha_{k_{v}, l_{v}})^{z_{ik_1}\times\cdots\times z_{ik_V} \times w_{j_{v}l_{v}}}.\\
\end{align*}

\subsection{Model Inference}\label{sec:s3s3}
Inference of the MVLBM aims to estimate the parameters $\bm{\Theta}$ that maximize the observed log-likelihood. While the EM algorithm \citep{dempsterMaximumLikelihoodIncomplete1977b} is a prominent method for performing estimation in the presence of latent variables, in the MVLBM context, difficulty in applying the EM algorithm arises due to the dependence structure among the features $x_{ij_{v}}$ in the model. In particular, it is computationally infeasible to estimate the expectation of the complete data log-likelihood, as this includes the term $\Pr(z_{ik_{1}} = 1,  \ldots z_{ik_{V}} = 1, w_{j_{v} l_{v}} = 1 | \bm{\Theta}^{(c)}, \bm{x})$. At each iteration of the EM algorithm, this would require, for each $v \in \lbrace 1, \ldots, V \rbrace $, $({K_{v}})^{n} \times ({L_{v}})^{d_{v}}$ calculations, making it computationally infeasible for even moderately-sized datasets.  

Several alternatives to EM exist for the LBM, such as the variational EM algorithm and the SEM-Gibbs algorithm. The variational EM approach imposes an additional assumption that the variational distribution $q(\bm{z}_1, \ldots, \bm{z}_V, \bm{w}_{1}, \ldots, \bm{w}_{V}| \bm{\Theta}, \bm{x})$ is restricted to be one where the hidden features are independent. Precisely, it is assumed that 
\begin{equation}
    q(\bm{z}_1, \ldots, \bm{z}_V, \bm{w}_{1}, \ldots, \bm{w}_{V}| \bm{\Theta}, \bm{x}) = p(\bm{z}_1| \bm{\Theta}, \bm{x})\cdots (\bm{z}_V| \bm{\Theta}, \bm{x})p(\bm{w}_{1}| \bm{\Theta}, \bm{x})\cdots p(\bm{w}_{V}| \bm{\Theta}, \bm{x}).
\end{equation} 
As the joint distribution is now factored with respect to the hidden features, estimating the joint cluster membership is replaced with estimating the row  and column cluster memberships independently. 

The variational approach is known to be susceptible to issues arising from poor initialization. To navigate these issues, we implement the modified approach introduced in \citep{keribinModelSelectionBinary2012}, integrating stochastic Gibbs sampling into the E-step of the model. The following five steps describe the $c$th iteration of the SEM-Gibbs algorithm. 
\begin{enumerate}
    \item \textit{Sampling row partitions} - At the $c$th iteration, a partition of the rows for each view is generated by sampling according to 
    \begin{equation*}
    p( z_{ik_{1}} = 1, \ldots, z_{ik_{V}}^{(c+1)} = 1 | \bm{x}, {\bm{w}_{1}}^{(c)}, \ldots, {\bm{w}_{V}}^{(c)}; {\bm{\alpha}_{1}}^{(c)}, \ldots, {\bm{\alpha}_{V}}^{(c)} ) \propto \pi_{k_{1}\ldots k_{V}}^{(c)} \prod_{v,j_v,l_{v}}
      p (x_{ij_{v}}; {\alpha_{k_{v} l_{v}}}^{(c)})^{w_{j_{v}l_{v}}}.
    \end{equation*}
    \item \textit{Row-wise M-Step} - This step proceeds by updating the co-cluster parameters to maximize the complete log-likelihood. This problem splits into $V + 1$ separate sub-problems, one for $\bm{\pi}$ and one for each set of cluster view parameters $\bm{\alpha}_{v}$. The $\bm{\pi}$ update has an analytical solution give by $\bm{\pi}^{(c + 1)}\in  \mathbb{R}^{K_{1} \times \cdots \times K_{V}}$ with 
\begin{equation*}
    \pi_{k_{1} \ldots k_{V}}^{(c+1)}  = \frac{1}{n}\sum_{i} z_{ik_{1}}^{(c+1)} \cdots z_{ik_{V}}^{(c+1)}.
\end{equation*}
The cluster parameters for the $v$th view are updated by solving a weighted maximum likelihood problem that is in exactly the same form as the M-step for a standard single-view LBM, making it easy to implement. As the parameters for the block clusters are updated again in the column-wise M-step, the parameter values at this stage are denoted $\bm{\alpha}_{v}^{(c +1/2)}$. The update is dependent on the type of features in the data. Section \ref{sec:s3s6} describes how to update the parameters for each feature type. 

\item \textit{Sampling column partitions} - For each view, the column partitions are sampled according to 
    \begin{multline*}
    p( {w_{j_{v}l_{v}}}^{(c+1)}  = 1 | \bm{x}, {\bm{z}_1}^{(c+1)}, \ldots, {\bm{z}_V}^{(c+1)}; {\bm{\alpha}_{1}}^{(c +1/2)}, \ldots, {\bm{\alpha}_{V}}^{(c+1/2)} )
        \propto \\ {\rho_{l_{v}}}^{(c)}\prod_{k_{1},\ldots, k_{V}} p (x_{ij_{v}}; {\alpha_{k_{v} l_{v}}}^{(c+1/2)})^{z_{ik_1}^{(c+1)}\times\cdots\times z_{ik_V}^{(c+1)}}.
    \end{multline*}
       \item \textit{Column-wise M-step} - The parameters to be found in the column-wise M-step are (1) the updates for the vectors ${\bm{\rho}_{v}}$ for the $(c + 1)$st iteration, and (2) the partial updates for the view-specific cluster parameters. 
The M-step for the columns proceeds separately for each view. For each view, the update of $\bm{\rho}_{v}$ has an analytical solution given by ${\bm{\rho}_{v}}^{(c+1)} \in \mathbb{R}^{L_{v}}$ with 
    \begin{equation*}
        {\rho_{l_{v}}}^{(c+1)} = \frac{1}{d_{v}} \sum_{j_{v} = 1}^{d_{v}} {w_{j_{v},l_{v}}}^{(c+1)}.
    \end{equation*}
    As this is the final update for the block cluster parameters, the values at this stage are denoted $\bm{\alpha}_{v}^{(c +1)}$. The view specific cluster parameters are updated in a similar way to the row-wise approach, and are explained in Section \ref{sec:s3s6}.
\item \textit{Missing values imputation} - For each view, samples with missing entries are imputed according to 
    \begin{multline*}
      p(\hat{x}_{ij_{v}}^{(c+1)} | \bm{x}, \bm{z}_1^{(c+1)}, \ldots,  \bm{z}_V^{(c+1)}, {\bm{w}_{1}}^{(c+1)}, \ldots, {\bm{w}_{V}}^{(c+1)}; \bm{\theta}^{(c+1)}) = \\\prod_{k_{1},\ldots ,k_{V}}\prod_{v,j_v,l_{v}} p({\hat{x}_{ij_{v}}}^{(c+1)} ; \alpha_{k_{v} l_{v}}^{(c+1)})^{z_{ik_1}^{(c+1)}\times\cdots\times z_{ik_V}^{(c+1)}\times{w_{j_{v}l_{v}}^{(c+1)}}}.
    \end{multline*}
\end{enumerate}

\subsection{Initialization \& Estimation}\label{sec:s3s4}
The SEM-Gibbs algorithm begins with an initialization of the partitions, provided by random assignment or using $k$-means++ clustering for continuous data. The mixing proportions of each block cluster and the block parameters are estimated from these initial partitions. The random initialization strategy can lead to empty clusters, particularly as the number of views and the number  of row clusters in each view grows. To address this issue, we implement the resampling strategy described in \citep{selosse_model-based_2020}. During the initial I iterations, equal to or fewer than the number of burn-in iterations, if a row or column cluster becomes empty, we sample a percentage of partitions in that view from a multinomial distribution with a common probability.

The SEM-Gibbs algorithm repeats these steps multiple times. The initial iterations are termed the burn-in period since the parameters have not stabilized yet. Consequently, only iterations occurring after the burn-in period are considered for parameter estimation. After completing the iterations post-burn-in, we combine them by taking the mean for continuous parameters and the mode for discrete parameters. This amalgamation yields a final estimation of the parameters for each block cluster. While SEM-Gibbs does not increase the log-likelihood at each iteration, the post-burn-in iterations produce an irreducible Markov chain with a unique stationary distribution expected to concentrate around the maximum likelihood parameter estimate.

\subsection{Model Selection}\label{sec:s3s5}
It is not feasible to use information criteria, such as the AIC and the BIC to select the optimal number of clusters, $K_{1}, \ldots, K_{V}, L_{1}, \ldots, L_{V}$, as they require evaluating the likelihood function at the maximum likelihood. Due to the dependency of the maximum likelihood on the observed data $\bm{x}$, this is not tractable for the MVLBM. A common alternative in the LBM literature is the approximation of the ICL criterion, often called the ICL-BIC, which relies on completed latent block information $\hat{\bm{z}}_1, \ldots, \hat{\bm{z}}_V, {\hat{\bm{w}}_{1}}, \ldots, {\hat{\bm{w}}_{V}}$. The criterion $ICL(K_1, \ldots, K_{V}, L_{1}, \ldots, L_{V})$ for the MVLBM model searches for the model with maximal completed log-likelihood. While exact computation of the ICL is not feasible for the MVLBM, the method of \citet{keribinEstimationSelectionLatent2015} yields an approximation as $n$ and $d_1, \ldots, d_V$ tend to infinity:
\begin{align*}
 ICL ~~ \approxeq ~~ &  \log p(\bm{x}, \hat{\bm{z}}_1, \ldots,\hat{\bm{z}}_V, \hat{\bm{w}}_{1}, \ldots, \hat{\bm{w}}_{V}; \bm{\Theta}) \\  & ~~~- \frac{\sum_{v} K_{v} -1}{2} \log n \\ & ~~~~~~- \sum_{v} \frac{L_{v} - 1}{2}\log d_{v} \\ & ~~~~~~~~~ -\sum_{v}\frac{\sum_{v} K_{v} L_{v} \eta_{v}}{2} \log (n \times d_{v}),
\end{align*}
where $\eta_{v}$ is the number of parameters for each block in the $v$th view, dependent on the feature type as described in Section \ref{sec:s3s6}. The derivation of this criterion is provided in  \ref{sec:a1}. It should be noted that exhaustively exploring the co-clustering for every possible combination of row and column cluster numbers $K_1, \ldots, K_V, L_1, \ldots, L_V$ is computationally impossible. In Section \ref{sec:s4}, we delve into a method that navigates the model space, guided by hypothesis tests assessing the null hypothesis of independence among row clusters in each view.


\subsection{Updates for Different Data Types}\label{sec:s3s6}
The MVLBM method is applicable to a broad range of data types. We provide the expressions for the distribution and the parameter updates used when computing the row and column M-steps in the SEM-Gibbs algorithm. We here implement MVLBM for nominal, ordinal, continuous, and count data. 

\subsubsection{Nominal Data}\label{sec:s3s6s1}
For a block $(k_{v}, l_{v}) $ of nominal data, we apply the multinomial distribution $\mathcal{M}(1, \bm{p}_{k_{v} l_{v}})$, where $\bm{p}_{k_{v} l_{v}}^{r} = (p_{k_{v} l_{v}}^{r})_{r = 1, \ldots, m}$ and $m$ is the number of levels taken by the categorical feature. The parameter of the block is $\bm{\alpha}_{k_{v}l_{v}} = \bm{p}_{k_{v} l_{v}}$ and the PMF of the block is 
\begin{equation*}
    f(x_{ij_{v}} | \bm{p}_{k_{v} l_{v}} ) = \prod_{r} \left(p_{k_{v} l_{v}}^{r}\right)^{\mathbb{I}(x_{ij_{v}} = r)},
\end{equation*}
where $\mathbb{I}(\cdot)$ is the identity function. The update of the parameter $p_{k_{v}l_{v}}^{r}$ is 
\begin{equation*}
    p_{k_{v} l_{v}}^{r} = \frac{1}{n_{k_{v} l_{v}}} \sum_{i, j_{v}} z_{i k_{v}} w_{j_{v}l_{v}} \mathbb{I}(x_{ij_{v}} = r),
\end{equation*}
where $n_{k_{v}l_{v}}$ is the number of elements in the block $(k_{v}, l_{v})$. The total number of parameters to be updated for the multinomial distribution for each view is $(m-1)K_{v}L_{v}$.

\subsubsection{Ordinal Data}\label{sec:s3s6s2}

In contrast to nominal data, conventionally modeled by the multinomial distribution, clustering ordinal data lacks a standardized distribution. In this study, we align with recent literature \citep{jacques_model-based_2018, selosse_analysing_2019, selosse_model-based_2020} and adopt the BOS distribution, as introduced \citep{biernackiModelbasedClusteringMultivariate2016}, to model ordinal data. The distribution has two parameters, a position parameter $\mu_{k_{v}l_{v}} \in \{1, \ldots, m\}$ and a precision parameter $\beta_{k_{v} l_{v}} \in \left[ 0, 1\right]$. Values of $\beta_{k_{v}l_{v}}$ closer to 1 imply the data is more concentrated about the value of the position parameter $\mu_{k_{v} l_{v}}$. The parameter of the block is $\bm{\alpha}_{k_{v} l_{v}} = (\mu_{k_{v} l_{v}}, \beta_{k_{v}l_{v}})$ and the PMF of the block is 
\begin{equation*}
    f(x_{ij_{v}} | \mu_{k_{v}l_{v}}, \beta_{k_{v} l_{v}} ) = \sum_{r = 0}^{m-1} C_r (\mu_{k_{v} l_{v}}, x_{ij_{v}})\times \left(\beta_{k_{v} l_v}\right)^{r},
\end{equation*}
where $C_r(\cdot)$ is a function that returns a constant dependent on $\mu$ and $\beta$. Inference of the parameters of the BOS distribution relies on an EM-algorithm. The details of the algorithm are contained in \citet{biernackiModelbasedClusteringMultivariate2016}. The total number of parameters to be updated for the BOS distribution for each view is $2K_{v}L_{v}$.

\subsubsection{Continuous Data}\label{sec:s3s6s3}
For a block $(k_{v}, l_{v}) $ of continuous data, we apply the univariate Gaussian distribution $\mathcal{N}(\mu_{k_{v} l_{v}}, \sigma_{k_{v} l_{v}}^2)$. The parameter of the block is $\bm{\alpha}_{k_{v} l_{v}} = (\mu_{k_{v} l_{v}}, \sigma_{k_{v} l_{v}}^2)$ and the PDF of the block is 
\begin{equation*}
    f(x_{ij_{v}} | \mu_{k_{v} l_{v}}, \sigma_{k_{v} l_{v}}^2) = \frac{1}{\sqrt{2 \pi \sigma_{k_{v} l_{v}}^2}} \exp \left( -\frac{(x_{ij_{v}} - \mu_{k_{v} l_{v}})^2}{2\sigma_{k_{v} l_{v}}^2} \right).
\end{equation*}
The update of the parameters is 
\begin{equation*}
    \mu_{k_{v}l_{v}} = \frac{1}{n_{k_{v} l_{v}}}\sum_{i, j_{v}} z_{i k_{v}} w_{j_{v} l_{v}} x_{ij_{v}},
\end{equation*}
and 
\begin{equation*}
    \sigma_{k_{v} l_{v}}^2 = \frac{1}{n_{k_{v} l_{v}}}\sum_{i, j_{v}} z_{i k_{v}} w_{j_{v} l_{v}} (x_{ij_{v}} - \mu_{k_{v} l_{v}})^2.
\end{equation*}
The total number of parameters to be updated for the Gaussian distribution for each view is $2K_{v}L_{v}$.

\subsubsection{Count Data}\label{sec:s3s6s4}
Count features are modelled by the Poisson distribution. To ensure the model is identifiable, we require the parameter of the Poisson distribution for a block $(k_{v}, l_{v}) $ to depend on marginals. We model this using the parameter $\lambda_{k_{v}l_{v}} = n_{i\cdot} n_{\cdot j_{v}} \delta_{k_{v} l_{v}}$, where $n_{i\cdot} = \sum_{j_{v}} x_{ij_{v}}$ and $n_{\cdot j_{v}} =  \sum_{i} x_{ij_{v}}$ are the number of instances in row $i$ and column $j_{v}$. As the parameters $n_{i\cdot}$ and $ n_{\cdot j_{v}}$ are independent of the clustering, the only parameter is $\bm{\alpha}_{k_{v} l_{v}} = \delta_{k_{v} l_{v}}$. The PMF of the block is 
\begin{equation*}
    f(x_{ij_{v}} | \delta_{k_{v} l_{v}} ) = \frac{1}{x_{ij_{v}}!} \exp(- n_{i\cdot} n_{\cdot j_{v}} \delta_{k_{v}l_{v}}) (n_{i\cdot} n_{\cdot j_{v}} \delta_{k_{v} l_{v}})^{x_{ij_{v}}}.
\end{equation*}
The update for the parameter of the Poisson distribution is 
\begin{equation*}
   \delta_{k_{v} l_{v}}  = \frac{1}{n_{k_{v}\cdot} n_{\cdot l_{v}} } \sum_{i, j_{v}} z_{i k_{v}} w_{j_v l_{v}} x_{ij_{v}},
\end{equation*}
where $n_{k_{v}\cdot} = \sum_{i, j_{v}} z_{ik_{v}} x_{ij_{v}}$ and $ n_{\cdot l_{v}}\sum_{i, j_{v}} w_{j_v l_{v}} x_{ij_{v}}$. The total number of parameters to be updated for the Poisson distribution for each view is $K_{v}L_{v}$.

\subsubsection{Mixed Type Data}\label{sec:s3s6s5}
For survey applications, each view may encompass different types of data. To this end, the MVLBM approach can be used with the mixed data LBM presented in \citep{selosse_model-based_2020}. Under this approach, the data matrix for each view consists of disjoint sets of features, each of different type. The estimation procedure of \citep{selosse_model-based_2020} will enforce the same row clustering across the sets of features within each view, and the MVLBM procedure will allow partial dependencies between the views as before. In this case, the number of row clusters in each view remains unchanged as $K_1, \ldots, K_V$, however, the number of column clusters within a view $v$ is a vector of values $(L_{v_1}, \ldots, L_{v_{S_v}})$, where $L_{v_s}$ is the number of column clusters in the $s$th set of the $v$th views and $L_v = \sum_{s = 1}^{S_v} L_{v_s}$. The column weight vector $\bm{\rho}_v$ is now a collection of vectors $(\bm{\rho}_{v_1}, \ldots, \bm{\rho}_{v_{S_v}})$, each capturing the disjoint clusters in the feature sets. Similarly, the number of parameters in the view $v$ is the sum of the number of parameters in each disjoint set, i.e., $\eta_v = \sum_s \eta_{v_s}$. In this context, the MVLBM method can be viewed as a generalization of the mixed data LBM method presented in \citep{selosse_model-based_2020}.

\section{Testing for Independence}\label{sec:s4}
Applying the MVLBM to multi-view data is costly, particularly for large datasets, where the potential number of row and column clusters in each view is high. In this case, the size of the model space of the MVLBM grows exponentially. It is also worth considering if the row clusterings of the data views provided by the LBM are related to each other, in advance of applying the MVLBM. To address this, we introduce a test for the null hypothesis, assessing the independence of row clusterings across two views within a single set of observations. This test is developed by adapting methodologies from classical mixture models \citep{gao_are_2020} and stochastic block models \citep{gao2022testing}.

\subsection{Model Specification}\label{sec:s4s1}
Considering the case with $V = 2$ views, i.e. $\bm{x} = \{\bm{x}_1, \bm{x}_2\}$. The marginal density of the row clusters within each view is a mixture density
\begin{align*}
    f(\bm{x}_{v} | \bm{z}_v) &= \sum_{\bm{w}_v} \prod_{j_v, l_v} \rho_{l_v}^{w_{j_vl_v}} \prod_{i, j_v,k_v, l_v}p(x_{ij_v} ; \alpha_{k_vl_v})^{z_{ik_v} w_{j_vl_v}}.
\end{align*}
In what follows, we denote this density as $\psi(\bm{x}_v ; \bm{\theta}_v)$. As before, we have 
\begin{equation*}
    p(\bm{z}_1, \bm{z}_2; \bm{\Theta}) = p(\bm{z}_1, \bm{z}_2; \bm{\pi}) = \prod_{i, k_1, k_2} \pi_{k_1k_2}^{z_{ik_1}z_{ik_2}},
\end{equation*}
where $\bm{\pi} \in \bm{\Delta}^{K_1 \times K_2}$. Further, suppose that each cluster has positive probability, namely $\sum_i \Pr(z_{ik_v} = 1) > 0 $ for every $k_v \in \{1, \ldots, K_v\}$.  The joint density of $\bm{x}$ is 
\begin{align*}
    f(\bm{x}; \bm{\Theta}) & = f(\bm{x}_1, \bm{x}_2; \bm{\pi}, \bm{\theta}_1, \bm{\theta}_2)\\
    &= \sum_{\bm{z}_1, \bm{z}_2} p(\bm{z}_1, \bm{z}_2; \bm{\pi}) f(\bm{x}_1, \bm{x}_2 | \bm{z}_1, \bm{z}_2)\\
    &= \sum_{\bm{z}_1, \bm{z}_2} p(\bm{z}_1, \bm{z}_2; \bm{\pi}) f(\bm{x}_1 | \bm{z}_1)f( \bm{x}_2 | \bm{z}_2)\\
    &= \sum_{\bm{z}_1, \bm{z}_2} \prod_{i, k_1, k_2} \pi_{k_1k_2}^{z_{ik_1}z_{ik_2}} \psi(\bm{x}_1 ; \bm{\theta}_1)\psi(\bm{x}_2 ; \bm{\theta}_2),
\end{align*}
where the third equality follows from the conditional independence of $\bm{x}_1$ and $\bm{x}_2$ given $\bm{z}_1$ and $\bm{z}_2$. 

Following \cite{gao_are_2020, gao2022testing}, it is useful to parameterize the matrix $\bm{\pi}$ in terms of a triplet ($\bm{\pi}_{1}, \bm{\pi}_{2}, \bm{C}$) that separates the within-view information from the between-view information. 
\begin{prop}[Proposition 1 of \citep{gao_are_2020}]\label{prop1}\label{prop:p1}
   Suppose $\bm{\pi}_1 \in \bm{\Delta}_{+}^{K_1}$ and $\bm{\pi}_2\in \bm{\Delta}_{+}^{K_2}$. Then, 
   \begin{equation*}
       \left\lbrace \bm{\pi} \in \bm{\Delta}^{K_1 \times K_2} : \bm{\pi} \bm{1}_{K_2} = \bm{\pi}_1, ~~ \bm{\pi}^T \bm{1}_{K_1} = \bm{\pi}_2 \right\rbrace = \left\lbrace \text{diag}(\bm{\pi}_1) \bm{C}\text{diag}(\bm{\pi}_2) : \bm{C} \in \mathcal{C}_{\bm{\pi}_1\bm{\pi}_2}   \right\rbrace, 
   \end{equation*}
   where $ \mathcal{C}_{\bm{\pi}_1\bm{\pi}_2} = \lbrace \bm{C} \in \mathbb{R}^{K_1 \times K_2} : C_{k_1k_2} \geq 0,~~ \bm{C} \bm{\pi}_{2} = \bm{1}_{K_1}, ~~ \bm{C}^T \bm{\pi}_{1} = \bm{1}_{K_2} \rbrace$.
\end{prop}

Proposition \ref{prop:p1} shows that any matrix $\bm{\pi}$ can be written as a product of its row sums, $\bm{\pi}_{1}$, its column sums, $\bm{\pi}_{2}$, and a matrix, $\bm{C}$. The joint density can be rewritten as 
\begin{align*}
    f(\bm{x}; \bm{\Theta}) & = f(\bm{x}_1, \bm{x}_2; \bm{\pi}, \bm{\theta}_1, \bm{\theta}_2) \\
    &= \sum_{(\bm{z}_1 \times \bm{z}_2) \in Z_1 \times Z_2} \prod_{i, k_1, k_2} (\pi_{k_1}C_{k_1k_2}\pi_{k_2})^{z_{ik_1}z_{ik_2}} \psi (\bm{x}_1 ; \bm{\theta}_1)\psi (\bm{x}_2 ; \bm{\theta}_2)\\
    &= f(\bm{x}_1, \bm{x}_2; \bm{\pi}_1, \bm{\pi}_2, \bm{C}, \bm{\theta}_1, \bm{\theta}_2). 
\end{align*}
The density of $\bm{x}_1$ and $\bm{x}_2$ is parameterized in terms of $\bm{\pi}_1, \bm{\pi}_2, \bm{C}, \bm{\theta}_1$, and $\bm{\theta}_2$. 
The marginal distributions of $\bm{x}_1$ and $\bm{x}_2$ can be expressed as 
\begin{align*}
    f(\bm{x}_v; \bm{\pi}_v, \bm{\theta}_v) 
    &= \sum_{\bm{z}_v} \prod_{i, k_v} \pi_{k_v}^{z_{ik_v}} \psi (\bm{x}_v ; \bm{\theta}_v)\\
    & = \sum_{(\bm{z}_v \times \bm{w}_v) \in Z_v \times W_v}\prod_{i, k_v} \pi_{k_v}^{z_{ik_v}}\prod_{j_v, l_v} \rho_{l_v}^{w_{j_vl_v}} \prod_{i, j_v,k_v, l_v}p(x_{ij_v} ; \alpha_{k_vl_v})^{z_{ik_v} w_{j_vl_v}},
\end{align*}
which is exactly the LBM for a single view dataset as defined previously. Note that the marginal distribution of $\bm{x}_1$ does not depend on $\bm{\pi}_2, \bm{\theta}_2$ or $\bm{C}$, and similarly the marginal distribution of $\bm{x}_2$ does not depend on $\bm{\pi}_1, \bm{\theta}_1$ or $\bm{C}$. The log-likelihood of the MVLBM model can thus be expressed as 
\begin{equation*}
    \ell(\bm{\pi}_1, \bm{\pi}_2, \bm{C}, \bm{\theta}_1, \bm{\theta}_2) = \sum_{i = 1}^{n} \log f(\bm{x}_{1_{i\cdot}}, \bm{x}_{2_{i\cdot}}; \bm{\pi}_1, \bm{\pi}_2, \bm{C}, \bm{\theta}_1, \bm{\theta}_2).
\end{equation*}

\subsection{Model Estimation}\label{sec:s4s2}
As discussed above, the SEM-Gibbs algorithm outlined in Section \ref{sec:s3s3} can be used to provide a stationary Markov chain of the parameter estimates with probability mass centred about the maximum likelihood estimates. Nevertheless, the computational expense associated with exhaustively exploring the entire model space may be prohibitive. A simpler option, proposed by \citep{gao_are_2020,gao2022testing}, notes that $\bm{\pi}_v$ and $\bm{\theta}_v$ can be estimated using the marginal likelihood for the $v$th view:
\begin{equation*}
    \ell(\bm{\pi}_v, \bm{\theta}_v) = \sum_{i = 1}^{n} \log f(\bm{x}_{v_{i\cdot}}; \bm{\pi}_v, \bm{\theta}_v).
\end{equation*}
The estimation for each individual data view can be completed separately, returning $\hat{\bm{\pi}}_1, \hat{\bm{\pi}}_2, \hat{\bm{\theta}}_1$ and $\hat{\bm{\theta}}_2$. Next, to estimate $\bm{C}$, the multi-view likelihood evaluated at $\hat{\bm{\pi}}_1, \hat{\bm{\pi}}_2, \hat{\bm{\theta}}_1$ and $\hat{\bm{\theta}}_2$ is maximized subject to the constraints of Proposition \ref{prop:p1}: 
\begin{equation*}
    \hat{\bm{C}} = \argmin_{\bm{C} \in \mathcal{C}_{\hat{\bm{\pi}}_1\hat{\bm{\pi}}_2}} \left[ - \ell(\hat{\bm{\pi}}_1, \hat{\bm{\pi}}_2, \bm{C}, \hat{\bm{\theta}}_1, \hat{\bm{\theta}}_2) \right],
\end{equation*}
where $ \mathcal{C}_{\hat{\bm{\pi}}_1\hat{\bm{\pi}}_2}= \lbrace \bm{C} \in \mathbb{R}^{K_1 \times K_2} : C_{k_1k_2} \geq 0,~~ \bm{C} \hat{\bm{\pi}}_{2} = \bm{1}_{K_1}, ~~ \bm{C}^T \hat{\bm{\pi}}_{1} = \bm{1}_{K_2} \rbrace$. \citet{gao_are_2020} show that this problem is convex, and develop an optimization strategy combining exponentiated gradient descent \citep{kivinen1997exponentiated} and the Sinkhorn-Knopp algorithm \cite{franklin1989scaling}. The algorithm for estimating the matrix $\hat{\bm{C}}$ is Algorithm 1 of \cite{gao_are_2020} and proceeds as follows: 

\begin{enumerate}
    \item Obtain the maximum likelihood estimates for the marginal parameters $\hat{\bm{\pi}}_1, \hat{\bm{\pi}}_2, \hat{\bm{\theta}}_1$ and $\hat{\bm{\theta}}_2$ using the SEM-Gibbs algorithm outlined above. 
    \item Define matrices $\bm{\hat{\psi}}_1 \in \mathbb{R}^{n \times K_1}$ and $\bm{\hat{\psi}}_2 \in \mathbb{R}^{n \times K_2}$ where 
    \begin{equation*}
        \hat{\psi}_{i k_1} = \psi_1(\bm{x}_{1_{i\cdot}}, \hat{\bm{\theta}}_1) ~~~ \text{and} ~~~ \psi_{i k_2} = \hat{\psi}_2(\bm{x}_{2_{i\cdot}}, \hat{\bm{\theta}}_2). 
    \end{equation*}
    \item Fix a step size $s > 0$ following the guidance provided in Theorem 5.3 of \citet{kivinen1997exponentiated}.
    \item Let $\hat{\bm{C}}^{(1)} = \bm{1}_{K_1} \bm{1}_{K_2}^T$. For $t = 1, \ldots $ until convergence:
        \begin{enumerate}
            \item Define $M_{k_1k_2} = \hat{C}^{(t)}_{k_1k_2} \exp\{s G_{k_1k_2} - 1\}$ where 
            \begin{equation*}
                G_{k_1k_2} = \sum_{i = 1}^{n} \frac{\hat{\psi}_{i k_1}\hat{\psi}_{i k_2}}{[\hat{\bm{\psi}}_{1_{i\cdot}}]^T \text{diag}( \hat{\bm{\pi}}_1) \hat{\bm{C}}^{(t)} \text{diag}( \hat{\bm{\pi}}_2)\hat{\bm{\psi}}_{2_{i\cdot}}},
            \end{equation*}
            and $\hat{\bm{\psi}}_{1_{i\cdot}}$ and $\hat{\bm{\psi}}_{2_{i\cdot}}$ are the $i$th rows of $\hat{\bm{\psi}}_{1}$ and $\hat{\bm{\psi}}_{2}$ respectively.
            \item Let $\bm{u}^{(0)} = \bm{1}_{K_2}$ and $\bm{v}^{(0)} = \bm{1}_{K_1}$. For $t' = 1, \ldots$ until convergence:
 \begin{equation*}
                    \bm{u}^{(t')} = \frac{\bm{1}_{K_2}}{\bm{M}^T \text{diag}(\hat{\bm{\pi}}_1) \bm{v}^{(t' -1)}}, ~~~ \bm{v}^{(t')} = \frac{\bm{1}_{K_1}}{\bm{M} \text{diag}(\hat{\bm{\pi}}_2 )\bm{u}^{(t')}},
                \end{equation*}
                where the fractions denote element-wise division. 

            \item Let $\bm{u}$ and $\bm{v}$ be the vectors to which $\bm{u}^{(t')}$ and $\bm{v}^{(t')}$ converge respectively. Let $\hat{\bm{C}}_{k_1k_2}^{(t + 1)} = u_{k_1} M_{k_1k_2} v_{k_2}$.
        \end{enumerate}
        \item Let $\hat{\bm{C}}$ denote the matrix to which $\hat{\bm{C}}^{(t)}$ converges, and let $\hat{\bm{\pi}} = \text{diag}(\hat{\bm{\pi}}_1) \hat{\bm{C}} \text{diag}(\hat{\bm{\pi}}_2)$.
\end{enumerate}

\subsection{Testing Independence}\label{sec:s4s3}
To motivate the application of the MVLBM, it is proposed to test the association between the row clusterings of the data views, that is to develop a test for the null hypothesis that $H_0 : \bm{C} = \bm{1}_{K_1} \bm{1}_{K_2}^T$, or equivalently that $H_0 : \bm{\pi} = \bm{\pi}_1 \bm{\pi}_{2}^T$. Following \citep{gao_are_2020, gao2022testing}, a likelihood ratio statistic is proposed to test $H_0$. 

The marginal maximum likelihood estimates, $\hat{\bm{\pi}}_1, \hat{\bm{\pi}}_2, \hat{\bm{\theta}}_1$ and $\hat{\bm{\theta}}_2$, are included in the test statistic:
\begin{align*}
    \log \tilde{\Lambda} &= \sup_{\bm{C} \in \mathcal{C}_{\hat{\bm{\pi}}_1\hat{\bm{\pi}}_2}} \ell(\hat{\bm{\pi}}_1, \hat{\bm{\pi}}_2, \bm{C}, \hat{\bm{\theta}}_1, \hat{\bm{\theta}}_2) - \ell(\hat{\bm{\pi}}_1, \hat{\bm{\pi}}_2, \bm{1}_{K_1} \bm{1}_{K_2}^T, \hat{\bm{\theta}}_1, \hat{\bm{\theta}}_2)  \\
    &= \ell(\hat{\bm{\pi}}_1, \hat{\bm{\pi}}_2, \hat{\bm{C}}, \hat{\bm{\theta}}_1, \hat{\bm{\theta}}_2) - [\ell(\hat{\bm{\pi}}_1, \hat{\bm{\theta}}_1) + \ell(\hat{\bm{\pi}}_2, \hat{\bm{\theta}}_2) ]\\
    & = \sum_{i = 1}^n \log \left[\frac{[\hat{\bm{\psi}}_{1_{i\cdot}}]^T \text{diag}( \hat{\bm{\pi}}_1) \hat{\bm{C}} \text{diag}( \hat{\bm{\pi}}_2)\hat{\bm{\psi}}_{2_{i\cdot}}}{[\hat{\bm{\psi}}_{1_{i\cdot}}]^T \hat{\bm{\pi}}_1\hat{\bm{\pi}}_2^T\hat{\bm{\psi}}_{2_{i\cdot}}} \right].
\end{align*}
To estimate the null distribution of $\log \tilde{\Lambda}$, it is proposed to take a permutation approach. Under $H_0$, the log-likelihood is identical under any permutation of the samples in each view. As such, repeated permutations of the samples of $\bm{x}_2$ are taken and the observed value of $\log \tilde{\Lambda}$ is compared to its empirical distribution in the permutation samples. Furthermore, the only necessary computation is a recalculation of the matrix $\bm{C}$ as the maximum liklehood estimates of the parameters $\hat{\bm{\pi}}_1, \hat{\bm{\pi}}_2, \hat{\bm{\theta}}_1$ and $\hat{\bm{\theta}}_2$ are invariant to the permutations. 

The permutation approach proceeds as follows:
\begin{enumerate}
    \item Compute $\log \tilde{\Lambda}$ as above using the data $\bm{x}_1$ and $\bm{x}_2$. 
    \item For $b = 1, \ldots, B$, where $B$ is the number of permutations:
        \begin{enumerate}
            \item Permute the observations in $\bm{x}_2$ to obtain $\bm{x}_2^b$
            \item Compute $\log \tilde{\Lambda}^b$ as above based on $\bm{x}_1$ and $\bm{x}_2^b$. 
        \end{enumerate}
        \item The p-value for testing $H_0: \bm{C} = \bm{1}_{K_1} \bm{1}_{K_2}^T$ is computed as $\frac{1}{B} \sum_{b=1}^B \mathbb{I}(\log \tilde{\Lambda} \leq \log \tilde{\Lambda}^b)$, where $\mathbb{I}(\cdot) = 1$ if the argument is true, and $0$ otherwise. 
\end{enumerate} 

\subsection{MVLBM Estimation}\label{sec:s4s4}
\begin{figure}[!ht]
    \centering
    \includegraphics[width=0.8\linewidth]{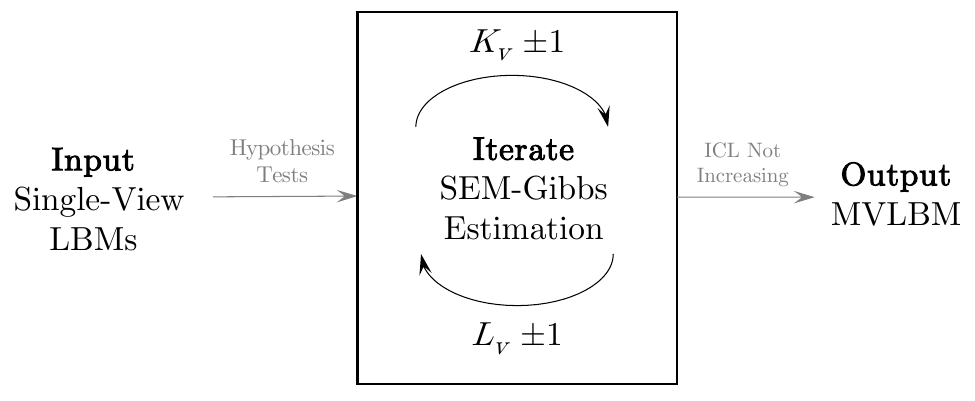}
    \caption{Given a set of single-view LBMs, we assess the dependence between the row clusters using a hypothesis testing procedure. The MVLBM estimation process begins with the single-view LBMs and, at each iteration, estimates the MVLBM with the number of row clusters $K_v$ and column clusters $L_v$ increased and decreased by 1. The number of clusters with the highest ICL is retained to initialize the next iteration. The process is repeated until the ICL stops increasing.}
    \label{fig:s4f6}
\end{figure}

It is proposed to integrate the results of the hypothesis testing framework into the estimation strategy for the MVLBM. Considering again a dataset $\bm{x}$ consisting of $V$ views, it is infeasible to exhaustively explore the entire model space for different numbers of row clusters and column clusters. As such, we first estimate single-view LBMs for each view $\bm{x}_v$ using the method of \citep{selosse_model-based_2020}. The hypothesis testing procedure is used to determine if a relationship between the views exists. As the MVLBM model places no requirements on the dependence between the latent row cluster memberships in each view, if the row clusters in view $v$ exhibit dependence with a least one other view in the dataset, it is retained for the MVLBM estimation. Beginning with the partitions from the single-view LBMs, we estimate the full MVLBM model using the SEM-Gibbs algorithm and compute the ICL for the model. To explore the model space, it is proposed to produce clusterings first increasing and then decreasing the number of row clusters in each view by 1, and subsequently increasing and then decreasing the number of column clusters in each view by 1. The clustering with the highest ICL is retained to initialize the next iteration. The process is terminated if the ICL does not improve and the final MVLBM is the clustering with the highest ICL value. The process is summarized in Figure \ref{fig:s4f6}.

\section{Sparsity Inducing Log Penalty}\label{sec:s5}
To mitigate the challenges associated with the exponential search space of MVLBM, we introduce a penalized likelihood approach for estimating the sparsity structure of $\bm{\pi}$, following the method introduced by \cite{huang_model_2017} and adapted for multi-view clustering in \cite{carmichael_learning_2020}. Applying standard sparsity-inducing penalties $\zeta(\cdot)$ on the entries of the cluster membership array $\bm{\pi}$, such as the Lasso, presents challenges due to the constraint that $\bm{\pi}$ must lie within the unit simplex. Moreover, exact zeros in $\bm{\pi}$ can create difficulties in enumerating the complete data log-likelihood. \cite{huang_model_2017} offers theoretical and experimental justification for the penalty $\zeta(\cdot) = \log( \delta + \cdot)$, where $\delta$ is a small positive value.

\begin{thm}[Theorem 3.1 of \cite{carmichael_learning_2020}]
Let $a_1, \ldots, a_K \geq 0$, $\sum_{k= 1}^K a_k = 1$, and $0 < \lambda < \frac{1}{K}$. Let $\bm{z}_{\delta} \in \mathbb{R}_{+}^{K}$ be a solution of the following problem for fixed $\delta > 0$, 
\begin{equation}
 \begin{aligned}\label{eq:s6e1}
    \min_{\bm{z}} \quad & - \sum_{k = 1}^K a_k \log(z_k) + \lambda \sum_{k = 1}^K \log(\delta + z_k) \\
    \text{s.t. } \quad &\bm{z} \geq 0 ~~ \text{and} ~~ \bm{z}^{T}\bm{1}_K = 1.
\end{aligned}   
\end{equation}
Then $\lim_{\delta \to 0} \bm{z}_{\delta} = \bm{z}_{0} \in \mathbb{R}^K$ where
\begin{equation}\label{eq:s6e2}
    z_{0_k} = \frac{(a_k - \lambda)_{+}}{\sum_{j = 1}^K (a_j - \lambda)_{+}}  ~~~ \text{for each } k \in 1, \ldots, K. 
\end{equation}
\end{thm}
The theorem shows that for small $\delta$, the global minimizer of (\ref{eq:s6e1}) is close to the normalized soft-threshold (\ref{eq:s6e2}). It is required that $\lambda < \frac{1}{K}$ to ensure that the denominator of (\ref{eq:s6e2}) is non-zero. 

For the MVLBM, the penalized likelihood criterion is:
\begin{equation}
    \max_{\bm{\Theta}} \quad \ell (\{\bm{x}_i\}_{i=1}^n | \bm{\Theta}) - \lambda\left(\sum_{k_1 = 1}^{K_1} \cdots \sum_{k_V = 1}^{K_V}  \log(\delta + \pi_{k_1, \ldots, k_v} )\right),
\end{equation}
where 
\begin{equation*}
    \ell(\{\bm{x}_i\}_{i=1}^n | \bm{\Theta}) = \sum_{i = 1}^n \log p(\bm{x}_i | \bm{\Theta}),
\end{equation*} is the observed data log-likelihood with $p(\bm{x}|\bm{\Theta})$ from (\ref{eq:s3eq4}) and $\delta >0$ is a small value. 

\subsection{Estimation}
\citet{huang_model_2017} and \citet{carmichael_learning_2020} employ a standard EM algorithm when applying a similar penalty to single- and multi-view mixture models, respectively. The penalty becomes operative in the M-step, where components with a prior probability below the threshold value $\lambda$ are deleted. Due to the considerations mentioned earlier, a modified EM-type algorithm, SEM-Gibbs, is necessary for MVLBM estimation. The only adjustment needed to the SEM-Gibbs formulation in Section 3 to incorporate the sparsity-inducing log penalty is in the Row-wise M-Step. The M-step update for the parameter $\bm{\pi}$ requires initially computing an array $\tilde{\bm{\pi}} \in \mathbb{R}^{K_{1} \times \cdots \times K_{V}}$
\begin{equation*}
    \tilde{\pi}_{k_{1} \ldots k_{V}}^{(c+1)}  = \frac{1}{n}\sum_{i} z_{ik_{1}}^{(c+1)} \cdots z_{ik_{V}}^{(c+1)}.
\end{equation*}
before applying the normalized soft-thresholding operator to delete components with prior probability less than the specified threshold:

\begin{equation*}
    \pi_{k_{1} \ldots k_{V}}^{(c+1)}  = \frac{( \tilde{\pi}_{k_{1} \ldots k_{V}}^{(c+1)} - \lambda)_{+}}{\sum_{k_{1}' = 1}^{K_1}\cdots\sum_{k_{V}' = 1}^{K_V} ( \tilde{\pi}_{k_{1}' \ldots k_{V}'}^{(c+1)} - \lambda)_{+}}.
\end{equation*}

Given the stochastic nature of the SEM-Gibbs algorithm and the challenge in initializing the MVLBM for inference, a common practice is to initialize the SEM-Gibbs algorithm without any penalty. Subsequently, the sparsity-inducing log penalty is applied for further runs of the SEM-Gibbs algorithm.

\section{Simulated Analysis}\label{sec:s6}
This section aims to demonstrate several key points: (1) the hypothesis testing procedure effectively detects dependence between clusterings when present; (2) the proposed SEM-Gibbs inference algorithm successfully uncovers the cluster structure of the data; (3) the model selection strategy is capable of selecting the correct number of clusters while sparsely exploring the model space; (4) the model accurately imputes missing values; and (5) the sparsity-inducing log penalty removes redundant clusters. The MVLBM method is implemented in Python and C++. The code to implement MVLBM and replicate the below experiments is available online.\footnote{\hyperlink{https://github.com/tobinjo96/MVLBM}{Github Repository: https://github.com/tobinjo96/MVLBM}} 

\subsection{Simulation Settings}
To assess the performance of the MVLBM method, we primarily consider two simulation settings. The first includes $V = 2$ views, each containing $n = 300$ observations and $S_v = 4$ sets of features with $d_1 = d_2 = d_3 = d_4 = 60$ features in each set. The second also includes $V = 2$ views, containing $n = 1200$ observations and $S_v = 4$ sets of features with $d_1 = d_2 = d_3 = d_4 = 300$ features in each set. 

The data for each view was generated from (\ref{eq:s3eq1} - \ref{eq:s3eq3}) with 
\begin{equation}\label{eq:s5eq1}
    \bm{\pi} = \frac{1 - \delta}{K^2} \bm{1}_K \bm{1}_K^T + \frac{\delta}{K} \bm{I}_K,
\end{equation}
for $K= 3$ and for a range of values of $\delta \in [0, 1]$. Setting $\delta = 0$ corresponds to independent clusterings across the views, and $\delta = 1$ corresponds to identical clusterings. Each view contains four data types: nominal ($m = 5$ levels), continuous, ordinal ($m = 3$ levels), and count data. The number of column clusters is $L_{v_s} = 3$ for each feature set $s$, within each view. The prior probability of each row cluster is uniform and equal to $1/3$. The prior probability of each column cluster is also uniform and equal to $1/12$. The parameters assigned to each block cluster are taken from \citep{selosse_model-based_2020} and are given in Table \ref{tab:s5t1}. 

\begin{table}
    \centering
    \caption{Configuration of the block cluster parameters for the simulated datasets.     \label{tab:s5t1}}
    \begin{tabular}{|p{1.2cm}|p{3.6cm}p{3.6cm}p{3.6cm}|}
    \hline
            & Nominal ($m = 5$) & & \\
         &$p_1, p_2, p_3, p_4, p_5$ & &  \\
             \cline{2-4}
         & Col C1 & Col C2 & Col C3\\   
   \hline
        Row C1 & 0.05,~0.05,~0.8,~0.05,~0.05 &0.1,~0.25,~0.3,~0.3,~0.05 &0.1,~0.2,~0.4,~0.2,~0.1 \\
        Row C2 & 0.05,~0.1,~0.7,~0.1,~0.05 & 0.8,~0.05,~0.05,~0.05,~0.05&0.4,~0.05,~0.1,~0.05,~0.4\\
        Row C3 & 0.2,~0.5,~0.2,~0.05,~0.05& 0.8,~0.05,~0.05,~0.05,~0.05& 0.05,~0.8,~0.05,~0.05,~0.05\\
         \hline
    \end{tabular}
\medskip

    \begin{tabular}{|p{1.2cm}|p{1.5cm}p{1.5cm}p{1.5cm}|p{1.5cm}p{1.5cm}p{1.5cm}|}
\cline{1-4} \cline{5-7}
        & Continuous & & & Ordinal& $(m = 3)$ & \\
         &$\mu, \sigma$ & &  & $\mu, \pi$& & \\
          \cline{2-4} \cline{5-7}
         & Col C1 & Col C2 & Col C3 & Col C1 & Col C2 & Col C3\\   
    \hline
        Row C1 & 100,~1&0.5,~5 &-90,~5 & 3,~0.4&1,~0.2 &3,~0.7 \\
        Row C2 & 10,~4& -15,~1&-95,~5&  2,~0.1& 3,~0.5&2,~0.8\\
        Row C3 & -20,~1&-30,~3 &500,~4&  2,~0.5&1,~0.8&2,~0.2\\
\cline{1-4} \cline{5-7}
    \end{tabular}\\
\medskip
    \begin{tabular}{|p{1.2cm}|p{1.5cm}p{1.5cm}p{1.5cm}|}
        \hline
        & Count & & \\
         &$\lambda$ & &  \\
         \cline{2-4}
         & Col C1 & Col C2 & Col C3\\   
    \hline
        Row C1 & 8.7&1.95 &8.16 \\
        Row C2 & 1.33& 1.95&25\\
        Row C3 & 7.27& 7.14&2.76\\
         \hline
    \end{tabular}

\end{table}

\begin{figure}[!ht]
    \centering
    \includegraphics[width=0.4\linewidth]{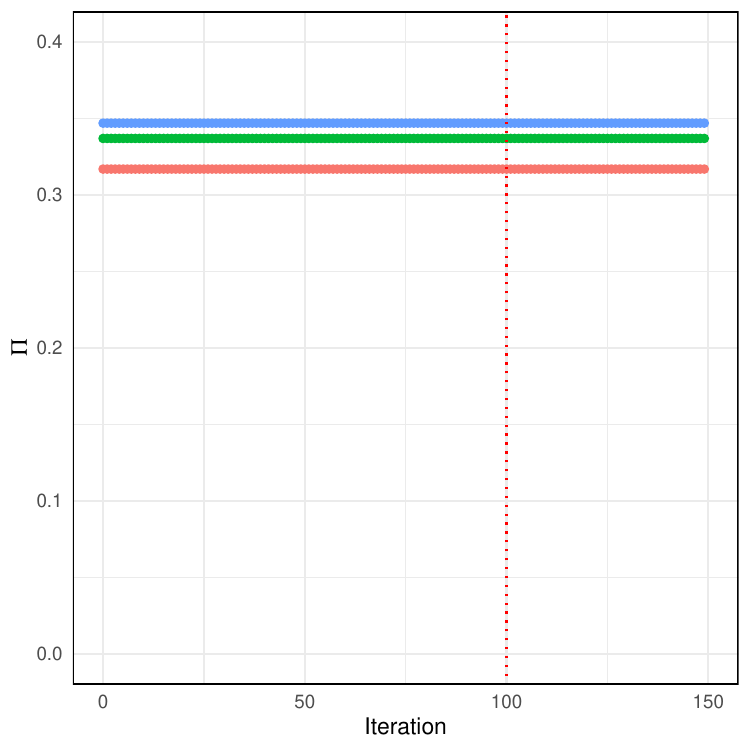}    \includegraphics[width=0.4\linewidth]{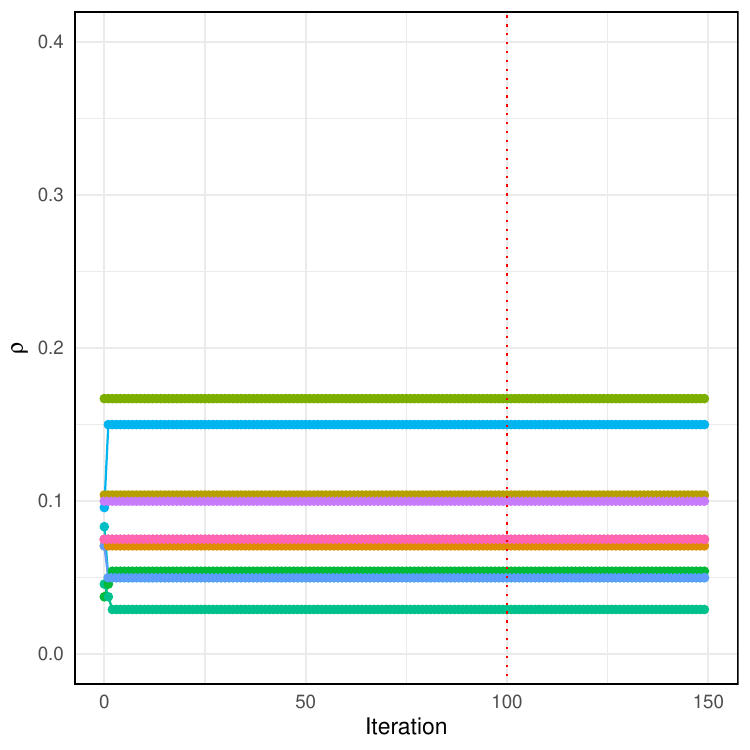}
    \caption{Parameter estimates for the row and column weights for an example dataset over 150 iterations of the SEM-Gibbs algorithm. The dashed red vertical line indicates the end of the burn-in period. }
    \label{fig:s5f1}
\end{figure}

For each setting, $20$ datasets are simulated. Algorithmic parameters are required for running the SEM-Gibbs algorithm for the single-view and multi-view data, as well as for the convex optimization algorithm used to estimate $\bm{\pi}$ for the hypothesis testing procedure. The total number of iterations for the SEM-Gibbs algorithm was set to 150, and the burn-in period was set to 100 iterations. 
The stability of the parameter estimates were assessed to judge if this number of iterations was sufficient. Sample trajectories of estimated parameters observed during the SEM-Gibbs algorithm are provided in Figure \ref{fig:s5f1}. 
In the case where a cluster becomes empty, it is proposed to resample a fraction of the assignments. We set the number of resampling iterations equal to 100, the same as the burn-in period, and fix the percentage of cluster assignments to be resampled at 20\%. For the convex optimization algorithm, the maximum number of iterations is set to 1000 for both the exponentiated gradient and Sinkhorn-Knopp algorithms. The step size $s$ is fixed at $1\times 10^{-5}$. 

\subsection{Hypothesis Testing}\label{sec:s5s2}
To investigate the Type I error and the power of the pseudo-likelihood ratio test of $H_0: \bm{C} = \bm{1}_{K_1} \bm{1}_{K_2}^T$, we apply it to the generated datasets at a nominal significance level of $\alpha = 0.05$. The method is assessed when the number of row clusters is correctly specified ($K_1 = K_2 = 3$) and when it is under-specified $(K_1 = K_2 = 2)$. The analysis is replicated for datasets with $n = 300$ and $n = 1200$. Furthermore, we compare the test to two competitor approaches. Following \citep{gao_are_2020}, the first competitor approach applies the $G$-test for independence, and the second approach applies a permutation test, comparing the adjusted rand index (ARI) of the estimated clusterings with clusterings produced by randomly permuting the data. We use $B = 200$ permutation samples to compute the p-value for each dataset for the permutation tests. 

\begin{figure}[!ht]
\centering
\includegraphics[width = 0.8\linewidth]{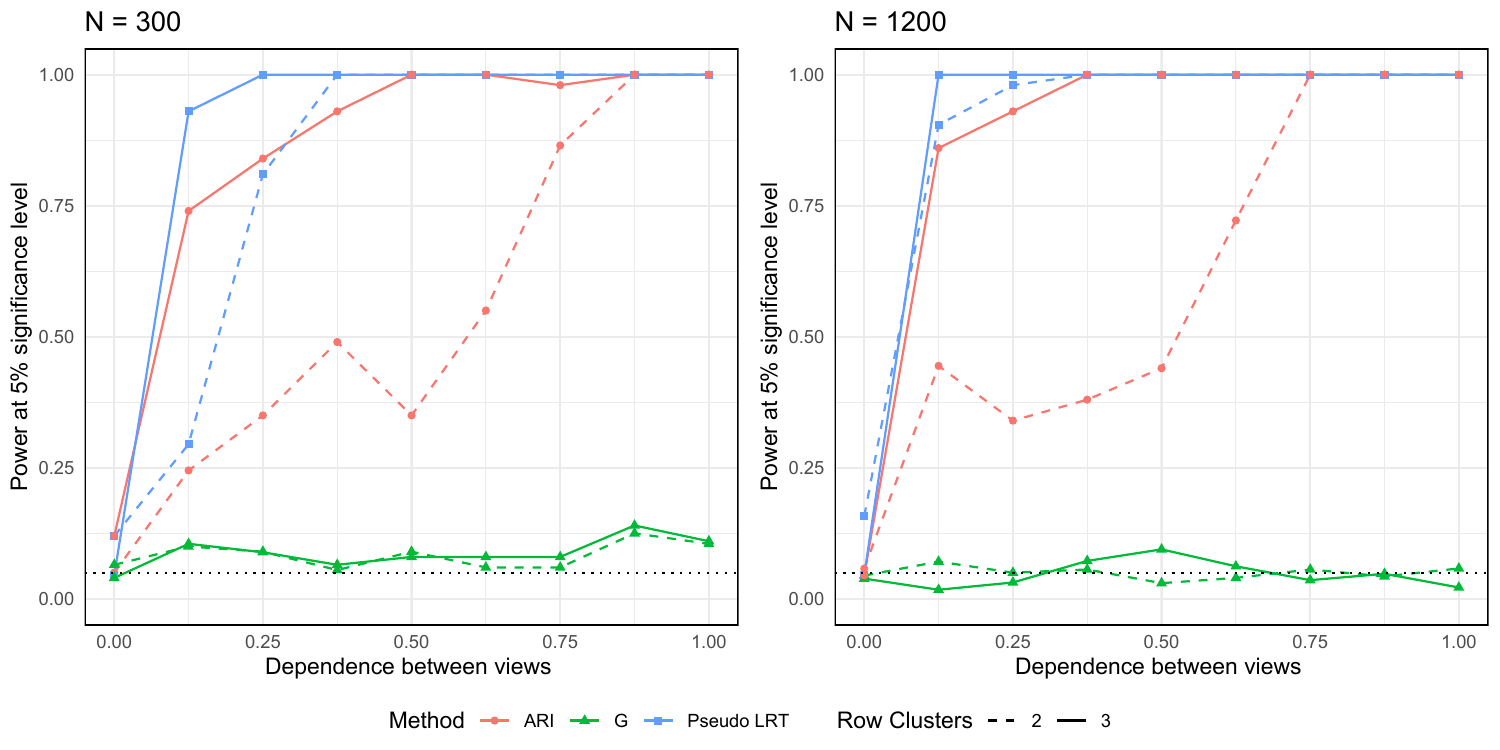}
\caption{\label{fig:s5f2} Power of the pseudo likelihood ratio test of $H_0: \bm{C} = \bm{1}_{K_1} \bm{1}_{K_2}^T$ for the simulation settings described in Section \ref{sec:s5s2}. The dependence between the views is $\delta$, defined in (\ref{eq:s5eq1}). The black dashed line at $y = 0.05$.  }
\end{figure}

The performance of the pseudo-likelihood ratio test, compared to its competitors, can be seen in Figure \ref{fig:s5f2}. The pseudo likelihood ratio test is seen to control the Type I error close to the nominal rate $\alpha = 0.05$, even when the number of clusters is misspecified. The power of the test increases dramatically with the dependence between the views, captured by $\delta$ in Equation (\ref{eq:s5eq1}) and greater power is exhibited when the number of data points increases. The performance of the test is not significantly inhibited when the number of clusters is misspecified, likely due to two `true' clusters being combined into one larger cluster. The G-test of independence is not seen to capture any meaningful information about the clusterings. While the ARI permutation test performs well when the dependence between the views is strong, it is less powerful than the pseudo likelihood ratio test in general.

\subsection{Parameter estimation}
\begin{figure}[!ht]
    \centering
    \includegraphics[width=0.4\linewidth]{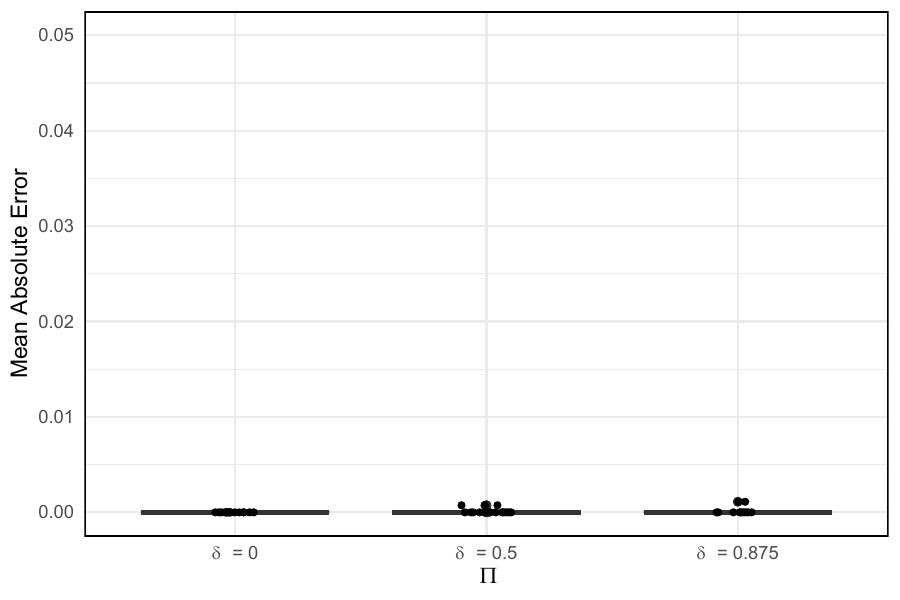}
    \includegraphics[width=0.4\linewidth]{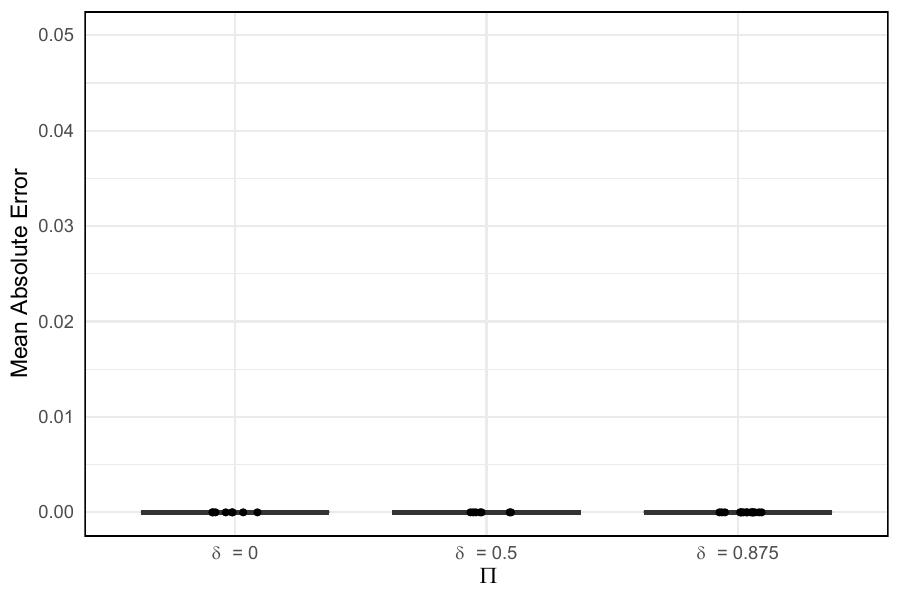}
    \caption{Mean absolute error for the row mixing proportions for $n = 300$ (left) and $n = 1200$ (right). }
    \label{fig:s5f3}
\end{figure}

\begin{figure}[!ht]
    \centering
    \includegraphics[width=0.7\linewidth]{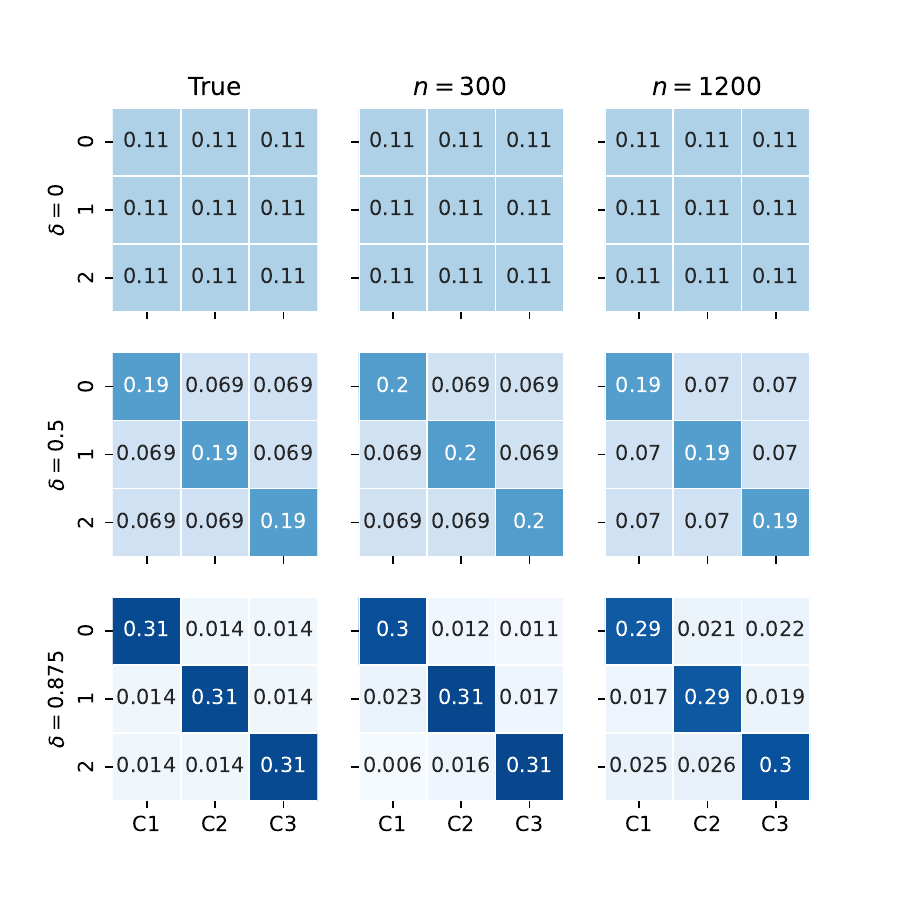}
    \caption{Estimates for the row mixing proportions $\bm{\pi}$ $n = 300$ (centre) and $n = 1200$ (right).  }
    \label{fig:s5f9}
\end{figure}

\begin{figure}[!ht]
    \centering
    \includegraphics[width=0.4\linewidth]{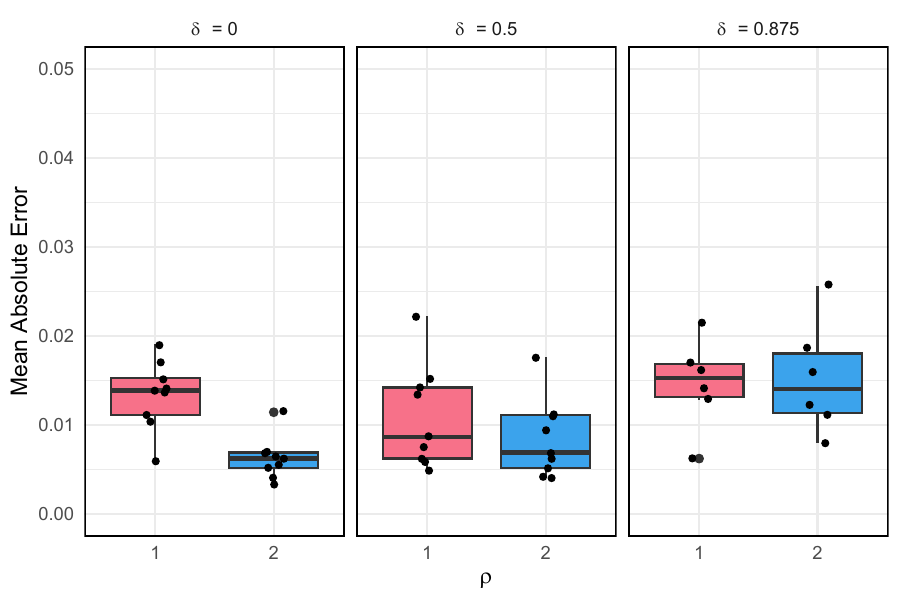}
    \includegraphics[width=0.4\linewidth]{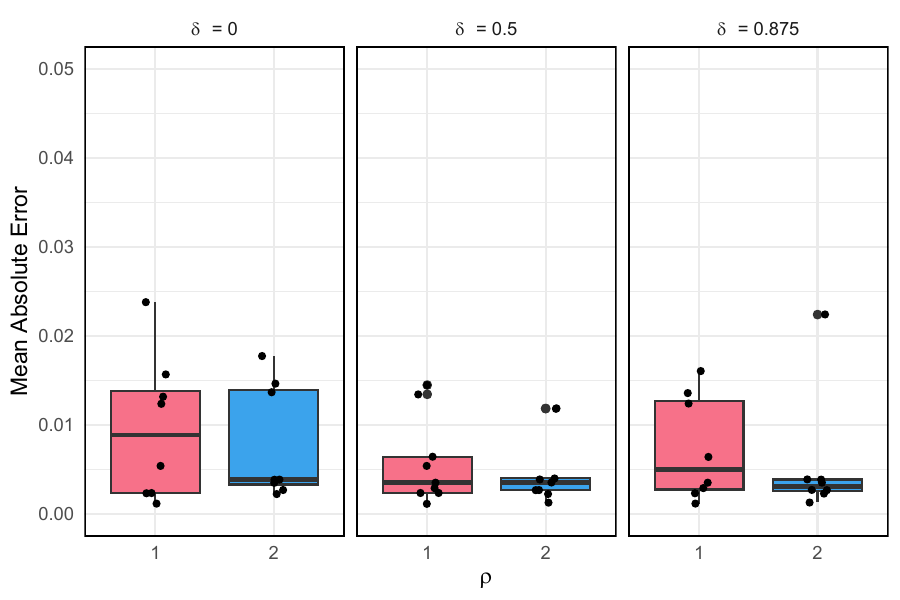}
    \caption{Mean absolute error for the column mixing proportions for $d_v = 60$ (left) and $d_v = 300$ (right). }
    \label{fig:s5f4}
\end{figure}

We next assess the ability of the SEM-Gibbs algorithm to perform accurate parameter estimation. The model was applied to 20 datasets for three configurations of $\bm{\pi}$, with $\delta = \{0, 0.5, 0.875\}$. We ensure $\delta < 1$ to avoid empty clusters in the estimation. The true number of row and column clusters was specified in each case. The mean absolute errors for the row proportions are provided in Figures \ref{fig:s5f3} and \ref{fig:s5f9} and the column proportions are provided in Figure \ref{fig:s5f4}. The mean absolute errors for each data type are provided in Table \ref{tab:s5t2} and Table \ref{tab:s5t3}. The ARI for the clusterings are provided in Table \ref{tab:s5t4}. For each data type, the mean absolute errors are low, indicating that the SEM-Gibbs procedure is capable of returning accurate estimates. 

We also assess the performance of the SEM-Gibbs algorithm in more challenging situations, producing datasets with less separation between the clusters. We change the true values of the parameters for the continuous features to those in Table \ref{tab:s5t5} retaining the three configurations of $\bm{\pi}$ used in the previous analysis. The ARI for the clusterings when the size of the datasets are $n = 300$ and $n = 1200$ are given in Table \ref{tab:s5t6}. We see that the SEM-Gibbs estimation algorithm remains capable of detecting the correct clustering, even when there is little separation between them in the continuous domain. 

Lastly, we assess the ability of the SEM-Gibbs algorithm to perform parameter estimation in the presence of missing data. New datasets were produced by randomly removing 15\% and 35\% of the data from each of the datasets used in the parameter estimation study. The ARIs for the clusterings are presented in Table \ref{tab:s5t7}. It is noted that the algorithm remains capable of detecting high quality clusterings in the presence of large amounts of missing data. 

\begin{table}[!ht]
    \centering
    \caption{Value of the block parameters' mean absolute error for datasets with $n= 300$, $d_{v_s} = 60$, over the three $\bm{\pi}$ parameters considered.  \label{tab:s5t2}}
    \small
    \begin{tabular}{|p{1cm}|p{1.8cm}p{1.8cm}p{1.8cm}p{1.8cm}|}
    \hline
         &Continuous &Nominal &Ordinal &Count  \\
         &$\mu, \sigma$ &$\bm{p}$ &$\mu, \pi$ &$\lambda \times 10^-5$\\
         \hline
         View 1 & 0.12, 0.00& 0.00&0.00, 0.03 & 0.00 \\
         View 2 & 0.11, 0.01& 0.00&0.00, 0.03 & 0.00 \\
         \hline
    \end{tabular}
\end{table}

\begin{table}[!ht]
    \centering
    \caption{Value of the block parameters mean absolute error for datasets with $n= 1200$, $d_{v_s} = 300$, over the three $\bm{\pi}$ parameters considered.\label{tab:s5t3}}
    \small
    \begin{tabular}{|p{1cm}|p{1.8cm}p{1.8cm}p{1.8cm}p{1.8cm}|}
    \hline
         &Continuous &Nominal &Ordinal &Count  \\
         &$\mu, \sigma$ &$\bm{p}$ &$\mu, \pi$ &$\lambda \times 10^-5$\\
         \hline
         View 1 &0.02, 0.00 &0.00 &0.01, 0.02 &0.00  \\
         View 2 &0.03, 0.00 &0.00 &0.02, 0.02 & 0.00 \\
         \hline
    \end{tabular}
\end{table}

\begin{table}[!ht]
    \centering
    \caption{Mean ARIs for the datasets with $n= 300$, $d_{v_s} = 60$ and $n= 1200$, $d_{v_s} = 300$.}
    \begin{tabular}{|p{1cm}|p{1cm}|p{1.5cm}p{1cm}p{1cm}p{1cm}|p{.1cm}|p{1.5cm}p{1cm}p{1cm}p{1cm}|}
    \hline
      & & View 1 & & & & & View 2 & & &    \\
            \cline{3-6} \cline{8-11}
       N   & Row & Continuous & Nominal & Ordinal & Count & & Continuous & Nominal & Ordinal & Count \\
       \hline
        300&1.00 &1.00 & 0.96&0.98 &1.00 & & 1.00&0.95&0.99 & 1.00   \\
      1200 &1.00 &1.00  &0.99 &0.98 &1.00 & &1.00 &0.99 &0.97 & 1.00    \\
       \hline
    \end{tabular}
    \label{tab:s5t4}
\end{table}

\begin{table}
    \centering
    \caption{Parameters for the blocks of continuous features for more challenging datasets.  \label{tab:s5t5}}
    \begin{tabular}{|p{1.2cm}|p{1.5cm}p{1.5cm}p{1.5cm}|}
    \hline
         &$\mu, \sigma$ & &  \\
          \cline{2-4} 
         & Col C1 & Col C2 & Col C3 \\   
    \hline
        Row C1 & 0.5, 1&0, 1 &0, 1 \\
        Row C2 & 0, 1& 0.5, 1&0, 1\\
        Row C3 & 0, 1&0, 1 &0.5, 1\\
         \hline
    \end{tabular}
\end{table}

\begin{table}[!ht]
    \centering
    \caption{Mean ARIs for the more challenging datasets.}    \label{tab:s5t6}
    \begin{tabular}{|p{1cm}|p{1cm}|p{1.5cm}p{1cm}p{1cm}p{1cm}|p{.1cm}|p{1.5cm}p{1cm}p{1cm}p{1cm}|}
    \hline
      & & View 1 & & & & & View 2 & & &    \\
            \cline{3-6} \cline{8-11}
       N   & Row & Continuous & Nominal & Ordinal & Count & & Continuous & Nominal & Ordinal & Count \\
       \hline
        300&1.00 &1.00 &0.93 & 0.97&1.00 & &1.00&0.92 &0.96 & 1.00   \\
       1200&1.00 &1.00  &0.99 &0.99 &1.00 & &1.00 &0.96 &0.98 & 1.00   \\
       \hline
    \end{tabular}
\end{table}

\begin{table}[!ht]
    \centering
    \caption{Mean ARIs for the datasets with missing values.}
    \begin{tabular}{|p{1cm}p{0.5cm}|p{0.6cm}|p{1.5cm}p{1cm}p{1cm}p{0.8cm}|p{.05cm}|p{1.5cm}p{1cm}p{1cm}p{0.8cm}|}
    \hline
    &  & & View 1 & & & & & View 2 & & &    \\
            \cline{4-7} \cline{9-12}
    \% NA &  N   & Row & Continuous & Nominal & Ordinal & Count & & Continuous & Nominal & Ordinal & Count \\
       \hline
     15 &300  & 0.96&0.97 &0.90 &0.91 &1.00 & & 1.00& 0.89& 0.92&1.00    \\
     35 & 300& 0.92& 0.89 &0.86 &0.83 &0.90 & &0.93 &0.86 &0.84 &0.77   \\
       \hline
   15 & 1200 & 1.00&0.96 & 0.95&1.00 &  1.00& & 0.94&0.94 & 0.95&1.00    \\
    35  &1200 & 0.93& 1.00 &0.93 &0.87 &0.95 & &1.00 &0.91 &0.88 &0.98  \\
\hline
    \end{tabular}
    \label{tab:s5t7}
\end{table}

\subsection{Model Selection}
In this section, the ability of the ICL criterion to select the correct number of row and column clusters is assessed. We also assess the model space search method introduced in Section \ref{sec:s4}. 
To To determine whether the ICL criterion is capable of selecting the correct number of row and column clusters, we complete an exhaustive search within a restricted model space. We restrict the analysis to datasets with $n = 300$. For each of the 20 datasets for one configuration of $\bm{\pi}$, with $\delta = 0.5$, we produce co-clusterings with $K_v \in \{2, 3, 4\}$ and $L_{v_s} \in \{2, 3, 4\}$. We restrict the number of clusters to be equal in each view, i.e. $K_1 = K_2$ and $L_{1_s} = L_{2_s}$ for $v \in \{1, 2\}$. This reduces the number of clusterings to be computed for each dataset from $3^{10}= 59049$ to $3^5 = 243$. The results, presented in Table \ref{tab:s5t8}, show that the ICL criterion regularly selects the clusterings with the correct number of clusters. 

\begin{table}[!ht]
    \centering
    \caption{Clusterings selected by the ICL criterion following an exhaustive search of the model space for 20 multi-view datasets.}
    \begin{tabular}{|c|cc c c |}
           \hline
      $n = 300$  & &  & & \\
       $(K_v, L_{v_1}, L_{v_2},L_{v_3},L_{v_4})$  &33333  & 33223&33222 &33322   \\
       \hline
       Counts   &  13& 5& 1& 1  \\
       \hline
    \end{tabular}
\label{tab:s5t8}
\end{table}

The proposed search method initially applies the greedy search method of \cite{selosse_model-based_2020} to explore the model space in each view. To assess the proposed search method, we produce co-clusterings for datasets with both $n = 300$ and $n = 1200$. Table \ref{tab:s5t9} presents the results from the single-view clusterings, highlighting the capability of the method described in \cite{selosse_model-based_2020} to detect the correct number of clusters. The results for the multi-view analysis are presented in Table \ref{tab:s5t10} in the form $(K_1, K_2, L_{1_1}, L_{1_2},L_{1_3},L_{1_4}, L_{2_1}, L_{2_2},L_{2_3},L_{2_4})$ capturing the number of row clusters and the number of column clusters for each of the four feature sets for the $v$th view respectively. When the model space is searched using the proposed method, having been initialized using the single-view clusterings, the ICL regularly selects clusterings with the correct number of components. 

\begin{table}[!ht]
    \centering
    \caption{Heuristic search on 40 single-view datasets using the method of \cite{selosse_model-based_2020}. }
    \begin{tabular}{|c|cc c|}
           \hline

      $n = 300$  & &  &\\

       $(K_v, L_{v_1}, L_{v_2},L_{v_3},L_{v_4})$  &33333  & 33323& 33223\\
       \hline
       Counts   &  24&13 &3\\
       \hline
    \end{tabular}
    \\
    \medskip
    \centering 
 \begin{tabular}{|c|cc |}
 \hline
$n = 1200$& &   \\

       $(K_v, L_{v_1}, L_{v_2},L_{v_3},L_{v_4})$  &33333  & 33223  \\
 \hline
       Counts   &  38& 2 \\
 \hline

    \end{tabular}
\label{tab:s5t9}
\end{table}

\begin{table}[!ht]
    \centering
    \caption{Heuristic search on 20 multi-view datasets initialized with the results of the single-view co-clusterings.}
    \begin{tabular}{|c|cc c c c | }
           \hline
      $n = 300$  & &  & & & \\
      \hline
          \begin{tabular}{@{}c@{}c@{}}       $(K_1, K_2)$ \\        $(L_{1_1}, L_{1_2},L_{1_3},L_{1_4})$ \\$( L_{2_1},L_{2_2},L_{2_3},L_{2_4})$\end{tabular}  &     \begin{tabular}{@{}c@{}c@{}}       $(3, 3)$ \\ $(3, 3, 3, 3)$ \\$(3, 3, 3, 3)$\end{tabular}& \begin{tabular}{@{}c@{}c@{}}       $(3, 3)$ \\        $(3, 2, 2, 3)$ \\$( 3, 2, 2, 3)$\end{tabular}& \begin{tabular}{@{}c@{}c@{}}       $(3, 3)$ \\        $(3, 3, 3, 3)$ \\$(3, 3, 3, 2)$\end{tabular}& \begin{tabular}{@{}c@{}c@{}}       $(3, 3)$ \\        $(3, 2, 2, 2)$ \\$(3, 2, 2,3)$\end{tabular} & \begin{tabular}{@{}c@{}c@{}}       $(3, 3)$ \\        $(3, 1, 2, 3)$ \\$(3, 2, 2,3)$\end{tabular}\\
       \hline
       Counts   & 13 &4 &1 & 1& 1\\
       \hline
           \end{tabular}\\
           
           \medskip
\centering
\begin{tabular}{|c|cc c |}
           \hline
$n = 1200$& & & \\
      \hline

    \begin{tabular}{@{}c@{}c@{}}       $(K_1, K_2)$ \\        $(L_{1_1}, L_{1_2},L_{1_3},L_{1_4})$ \\$( L_{2_1},L_{2_2},L_{2_3},L_{2_4})$\end{tabular}  &     \begin{tabular}{@{}c@{}c@{}}       $(3, 3)$ \\ $(3, 3, 3, 3)$ \\$(3, 3, 3, 3)$\end{tabular}& \begin{tabular}{@{}c@{}c@{}}       $(3, 3)$ \\        $(3, 2, 2, 3)$ \\$( 3, 2, 2, 3)$\end{tabular}& \begin{tabular}{@{}c@{}c@{}}       $(3, 3)$ \\        $(3, 3, 3, 3)$ \\$(3, 3, 3, 2)$\end{tabular} \\
 \hline
       Counts   & 15 &3 &2 \\
 \hline

    \end{tabular}
\label{tab:s5t10}
\end{table}

\subsection{Imputation Performance}
\begin{figure}[!ht]
    \centering
    \includegraphics[width=0.48\linewidth]{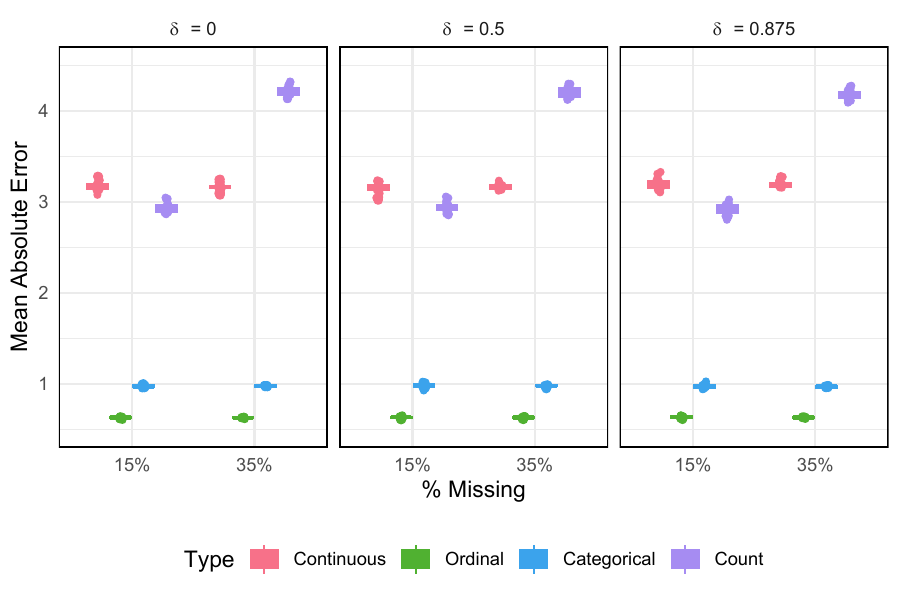}
    \includegraphics[width=0.48\linewidth]{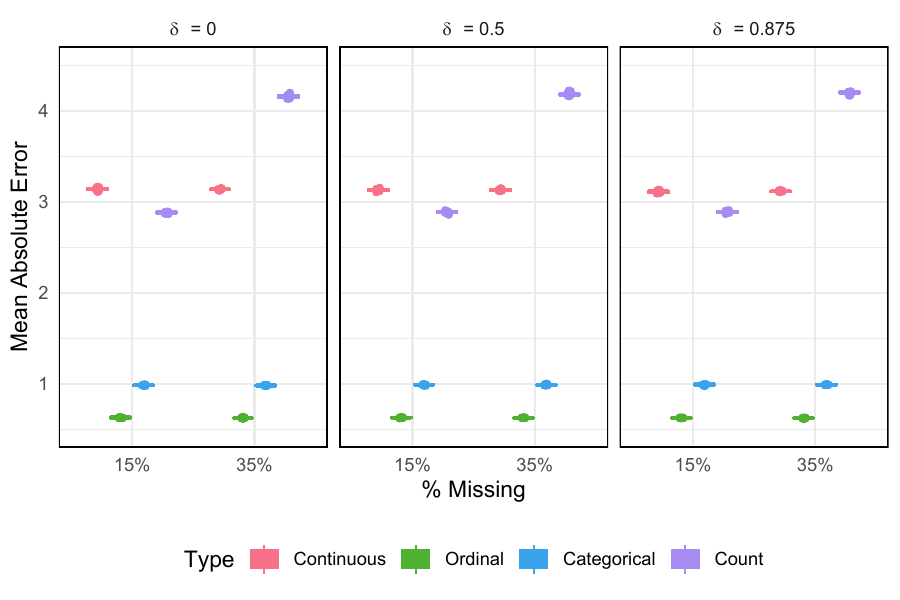}
    \caption{Mean absolute error for the imputed data for $n = 300$ (left) and $n = 1200$ (right). }
    \label{fig:s5f11}
\end{figure}

To assess the ability of the SEM-Gibbs algorithm to accurately impute missing data, it is applied to datasets with $\delta = \{0, 0.5, 0.875\}$ and with 15\% and 35\% of the values removed. Following training, the missing values are imputed using their cluster positions and the mean absolute error is computed and presented in Figure \ref{fig:s5f11}. The SEM-Gibbs algorithm is able to accurately impute missing values in the data. The scale of the imputation error also depends on the inherent variation within the clusters. This is evident in the outsized error experienced by continuous and count data compared to other data types in the study.  It is noteworthy that the mean absolute error for the imputed data remains independent of both the fraction of missing data points and the strength of dependency between the views. The mean absolute error for the two data sizes, $n = 300$ and $n = 1200$, are also similar, with the smaller dataset exhibiting only slightly more variation in the estimation error. 

\subsection{Sparsity Penalty}
The ability of the logarithmic penalty to induce sparsity in the clusterings is tested using the datasets with $\delta = 1$. For each of these datasets, we assess the performance of the penalty to remove redundant clusters from the joint space. We apply the penalty with values of $\lambda = ({i/10}^2)/{3^2}$ for $i \in \{0, 1, \ldots, 10\}$ to assess the sensitivity of the results to the value of $\lambda$. It is required to ensure that $\lambda < 1/3^2$, the reciprocal of the product of the row clusters in each view. The datasets with 15\% and 35\% of the values missing are also included in the analysis. The penalty value $\lambda$ was set to 0 during the burn-in period of the SEM-Gibbs algorithm to avoid undesired pruning due to the variation at early iterations. 
\begin{figure}[!ht]
    \centering
    \includegraphics[width=0.7\linewidth]{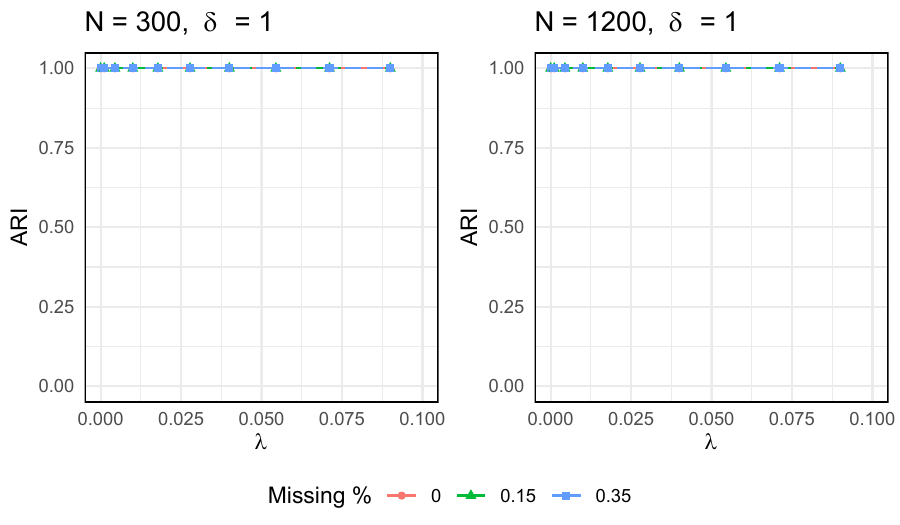}
    \caption{Adjusted Rand Index for the row clusterings returned by the MVLBM with the sparsity inducing log penalty. }
    \label{fig:s5f12}
\end{figure}

\begin{figure}[!ht]
    \centering
    \includegraphics[width=0.7\linewidth]{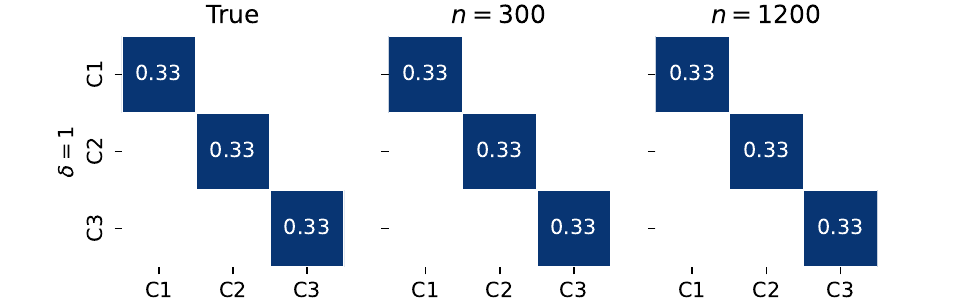}
    \caption{Estimates for the row mixing proportions $\bm{\pi}$ $n = 300$ (centre) and $n = 1200$ (right) for the sparsity induced clusterings.}
    \label{fig:s5f13}
\end{figure}

As the penalty influences only the row clusters returned by the MVLBM, we present the ARIs of the row clusterings for different values of $\lambda$ in Figure \ref{fig:s5f12}. The MVLBM returns a perfect clustering for every value of $\lambda$ and is not influenced by missing data. The results are supported by the estimates for the row mixing proportions $\bm{\pi}$, presented in Figure \ref{fig:s5f13}.

\section{Co-Clustering Chronic Lymphocytic Leukaemia Data}\label{sec:s7}
\begin{figure}[!ht]
    \centering
    \includegraphics[width=0.9\linewidth]{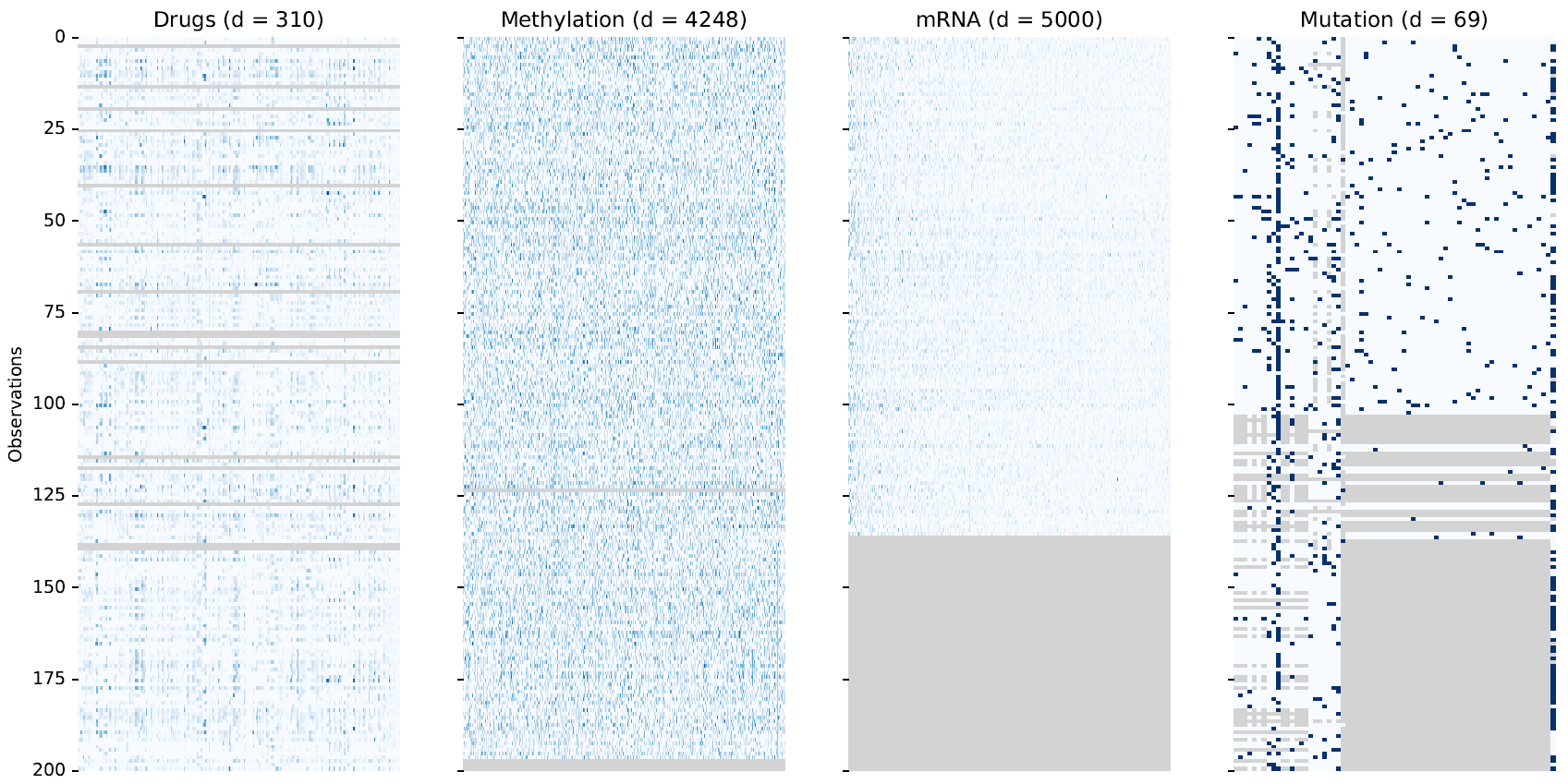}
    \caption{Overview of the CLL data views. The views are shown in different columns with darker values indicating a larger observation. Missing samples are shown in grey.}
    \label{fig:s6f1}
\end{figure}

In this section, we apply the MVLBM to a multi-view Chronic Lymphocytic Leukaemia (CLL) dataset, which combines drug response measurements (Drugs) with DNA methylation assays (Methylation), transcriptome profiling (mRNA), and somatic mutation status (Mutation), originally collected by \cite{dietrich2018drug}. This dataset, previously utilized to showcase the capabilities of multi-view factor analysis methods \cite{argelaguet2018multi}, focuses on the somatic mutation status of the immunoglobulin heavy-chain variables region (IGHV) as the primary quantity of interest for sample participants. This biomarker, impacting clinical care significantly, is typically considered binary in practice. A graphical summary of the data views are shown in Figure \ref{fig:s6f1}.


The estimation procedure followed the outlined steps, initially co-clustering each data view independently. We employed 20 random initializations for each view. Given the availability of two categories for IGHV status, we constrained the number of row clusters in each view to $K_v = 1$. Initially, for each view, the number of column clusters was set to $L_v = 1$. The growth in the number of column clusters was conducted greedily, utilizing the method proposed by \cite{selosse_model-based_2020}. After completing the iterative search of the model space, the model with the highest value of the ICL criterion was selected. The resulting number of row and column clusters for each view is presented in Table \ref{tab:s6t1}.

\begin{table}[!ht]
    \centering
    \caption{Number of row and column clusters returned from (1) the single-view and (2) the multi-view co-clustering of the CLL data views. }
    \begin{tabular}{|p{2cm}|p{1.5cm} p{1cm} |p{0.1cm}| p{1.5cm}p{1cm}|}
    \hline
    & Single & & & Multi & \\
    
    View & $K_v$ & $L_v$ & & $K_v$ & $L_v$\\
    \hline
       Drugs  & 2 & 7 & & 2& 7\\
       Methylation  &2 & 10& & 2& 10\\
       mRNA & 2& 11& & 2& 11\\
       Mutation & 2& 4& & 2& 3\\
    \hline
    \end{tabular}
    \label{tab:s6t1}
\end{table}

\begin{figure}[!ht]
    \centering
    \includegraphics[width=0.32\linewidth]{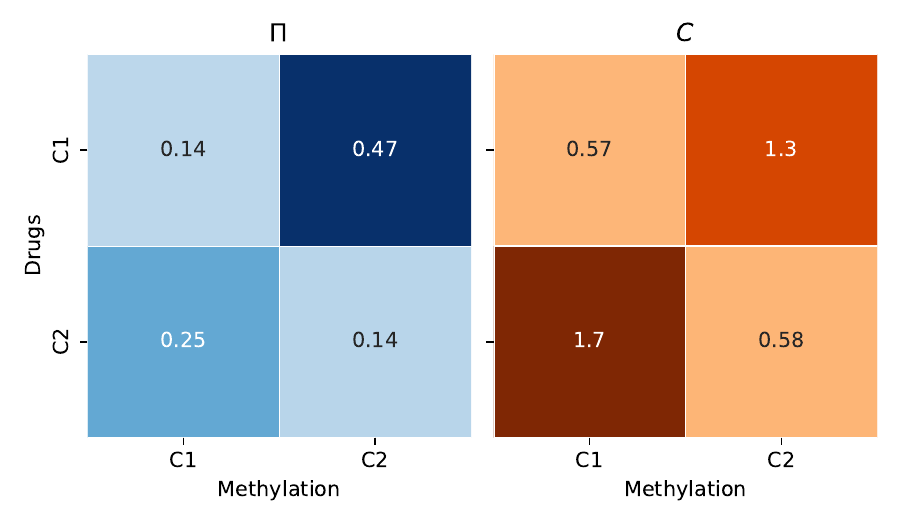}
    \includegraphics[width=0.32\linewidth]{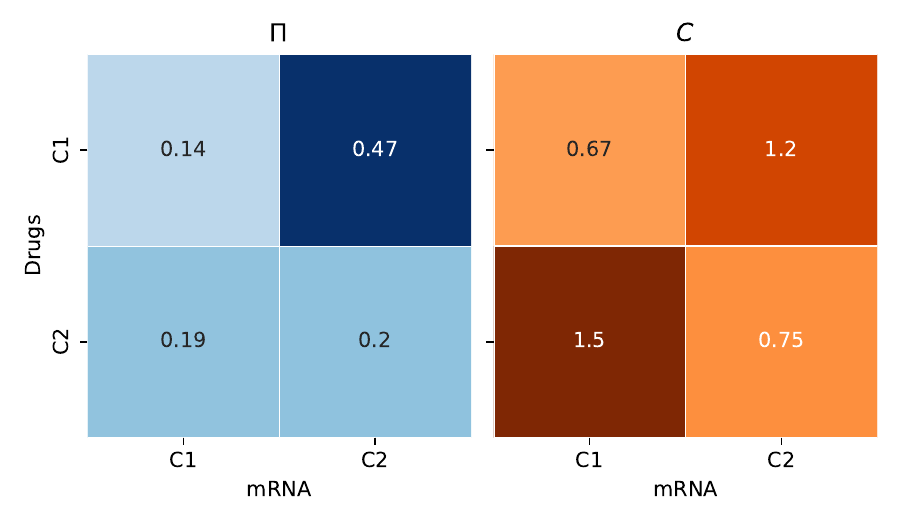}
    \includegraphics[width=0.32\linewidth]{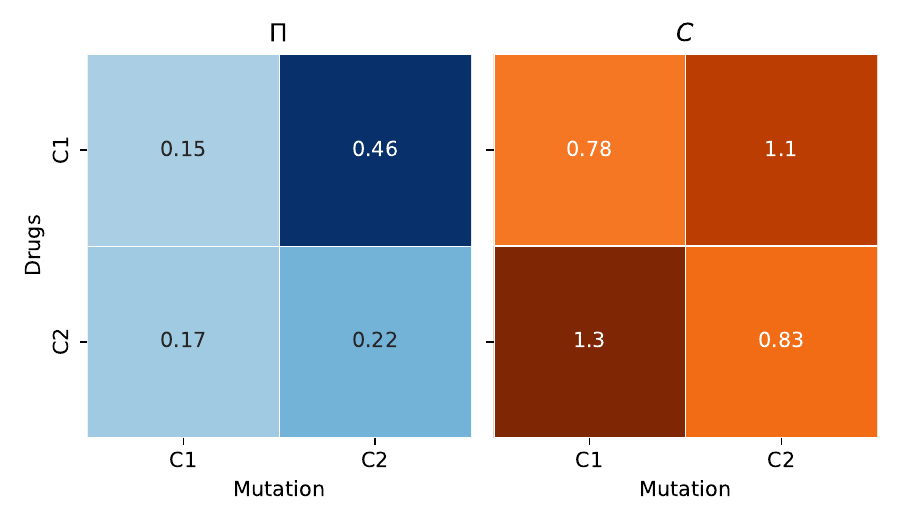}
    \includegraphics[width=0.32\linewidth]{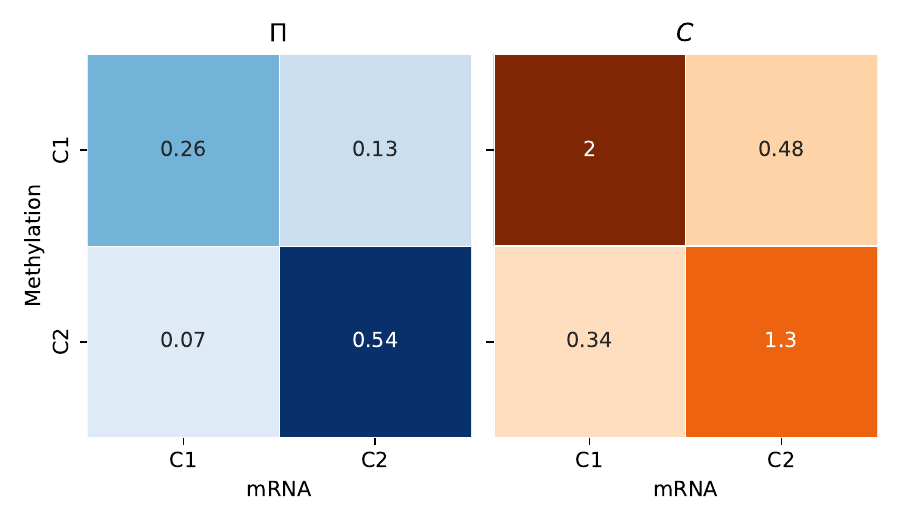}
    \includegraphics[width=0.32\linewidth]{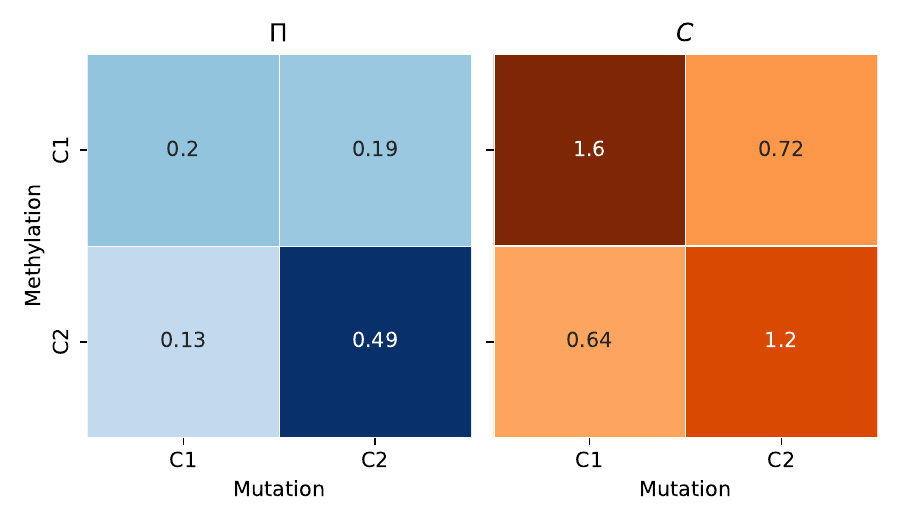}
    \includegraphics[width=0.32\linewidth]{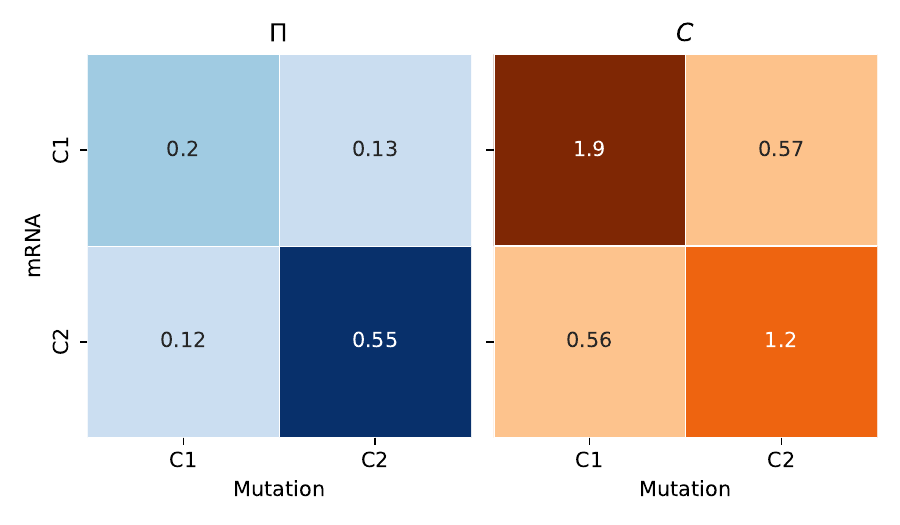}
    \caption{Pairwise marginal views of $\bm{\pi}$ (blue) and $\bm{C}$ (orange) for the CLL data views estimated using the convex estimation procedure of Section \ref{sec:s4}. }
    \label{fig:s6f2}
\end{figure}

Prior to applying the MVLBM, a pairwise hypothesis testing procedure was conducted on the row clusterings of the data views. The corresponding p-values are displayed in Table \ref{tab:s6t2}. Notably, all row clusterings were found to have significant relationships. Figure \ref{fig:s6f2} presents the estimated $\bm{\pi}$ and $\bm{C}$ matrices for each pair of views.

\begin{table}[!ht]
    \centering
    \caption{P-values for the test of no association between the row clusterings of the CLL data views. Significant results are highlighted in bold. }
    \begin{tabular}{|p{2cm}|p{1.5cm}p{2cm}p{1.72cm}|}
    \hline
         &  Drugs & Methylation & mRNA \\
         \hline
        Methylation & \textbf{0.000}& -& -\\
        mRNA & \textbf{0.000}& \textbf{0.000}&- \\
        Mutation & \textbf{0.000}& \textbf{0.000}& \textbf{0.000}\\
        \hline
    \end{tabular}
    \label{tab:s6t2}
\end{table}

Having detected the significant relationships between the row clusters collected in each view, we proceed by fitting an MVLBM to the data. The model is initialized using both the single-view co-clustering results and random initialization. Furthermore, we apply the sparsity inducing log penalty with values of $\lambda = {i/10}^2/{2^4}$ for $i \in \{0, 1, \ldots, 10\}$. The penalties are restricted to be less than $1/2^4$, 
the reciprocal of the number of joint space row clusters. As for the single-view clusterings, the number of row clusters is fixed at $K_v = 2$ for each view, and the number of column clusters is explored in the greedy manner described above. The optimal model is chosen using the ICL criterion. 

\begin{figure}[!ht]
    \centering
    \includegraphics[width=0.32\linewidth]{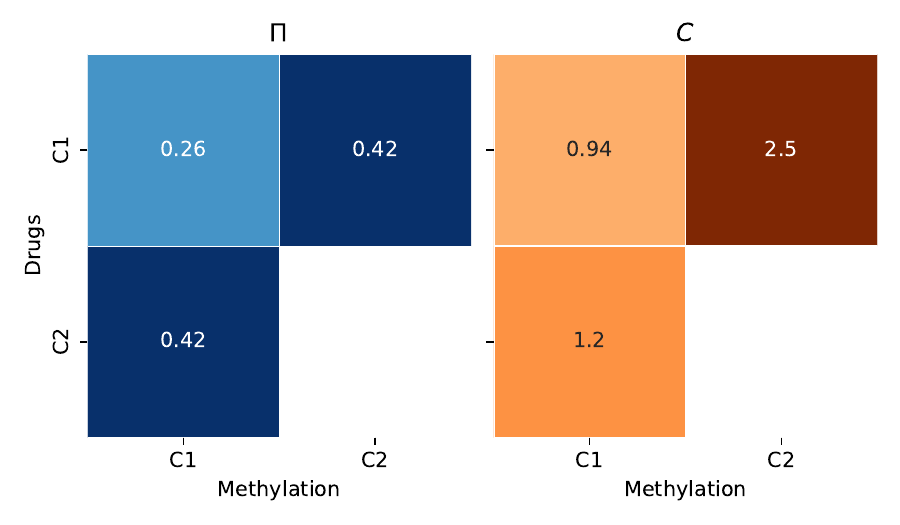}
    \includegraphics[width=0.32\linewidth]{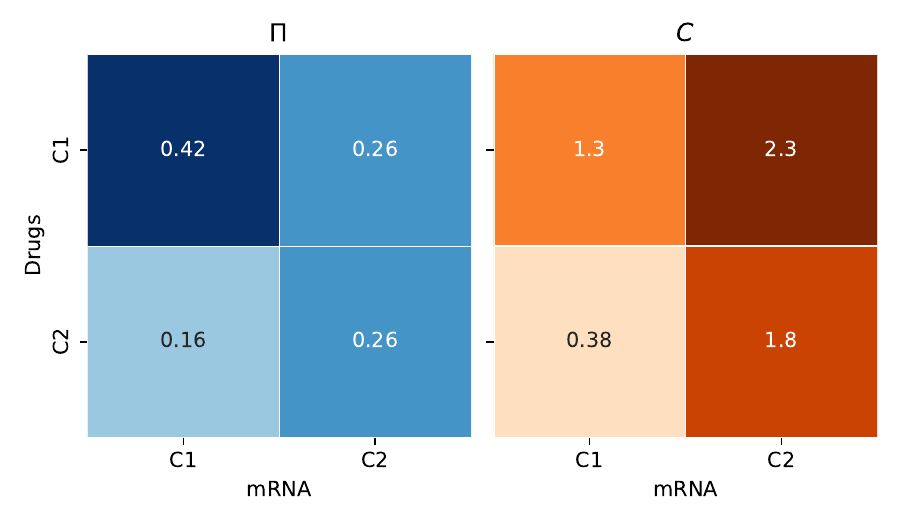}
    \includegraphics[width=0.32\linewidth]{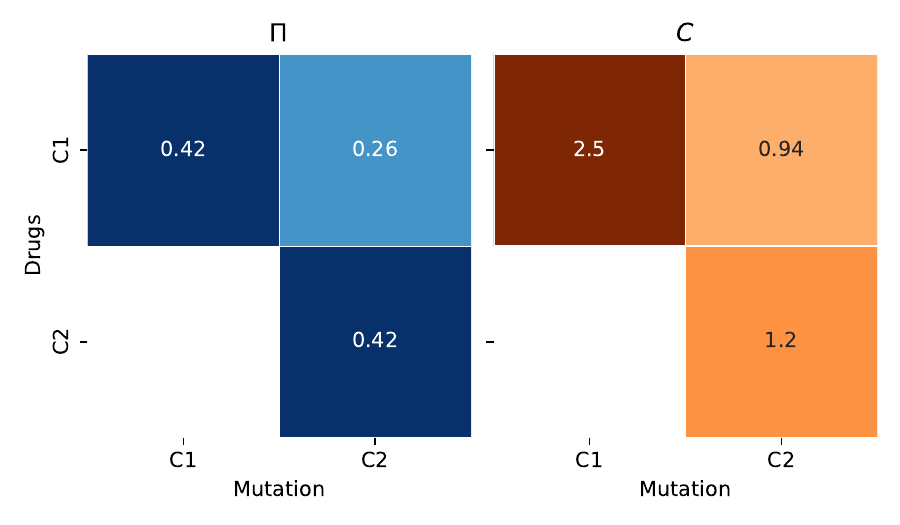}
    \includegraphics[width=0.32\linewidth]{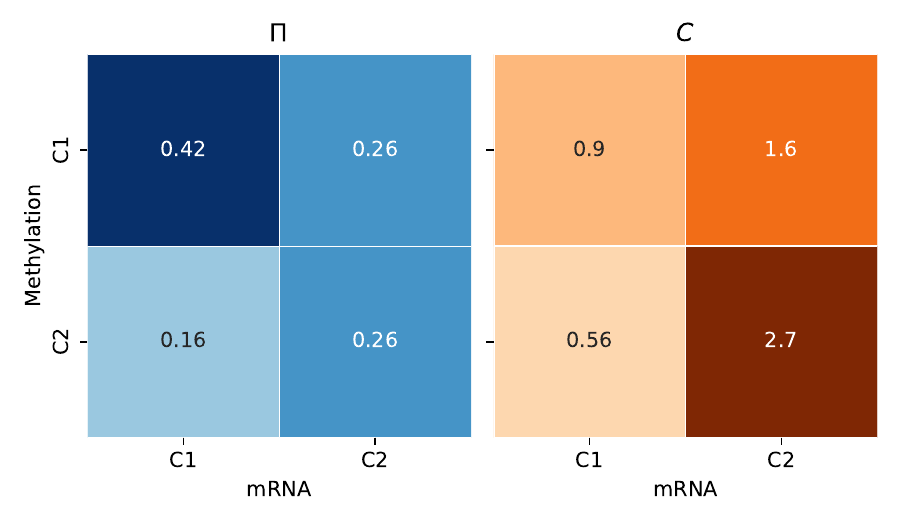}
    \includegraphics[width=0.32\linewidth]{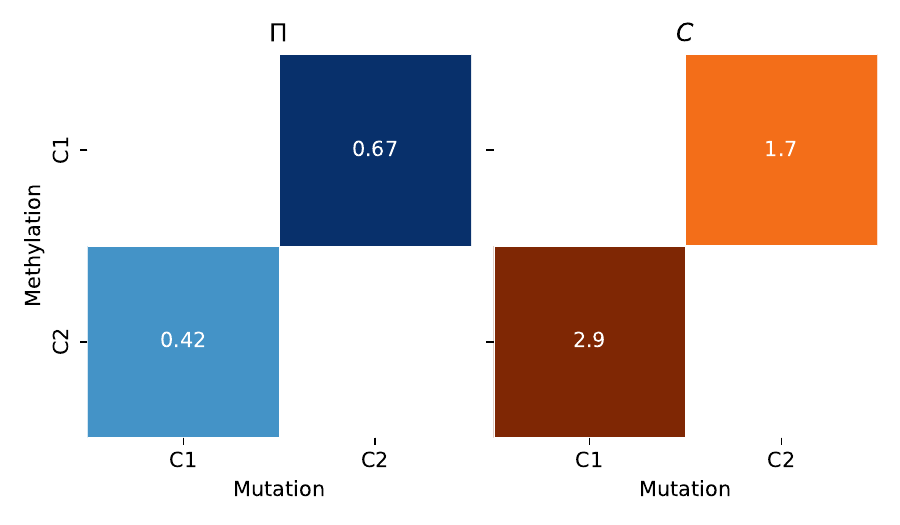}
    \includegraphics[width=0.32\linewidth]{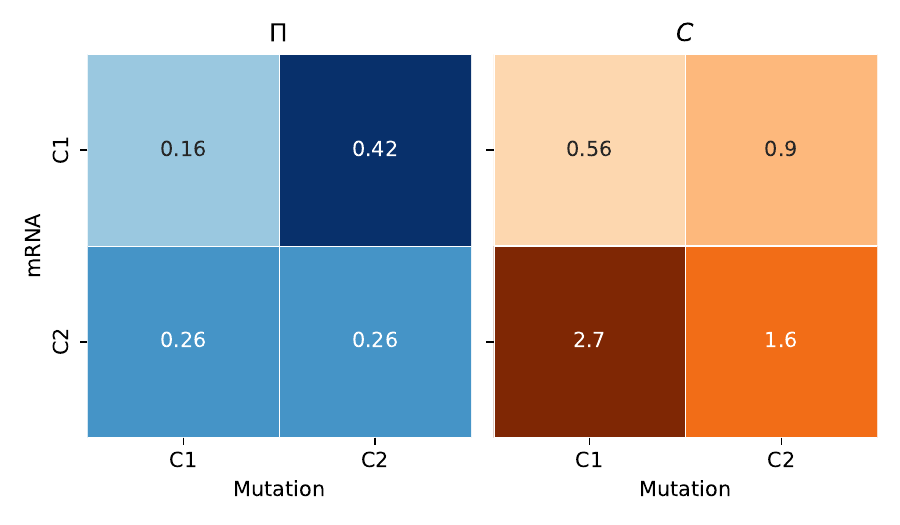}
    \caption{Pairwise marginal views of $\bm{\pi}$ (blue) and $\bm{C}$ (orange) for the CLL data views estimated by the log penalized MVLBM with $\lambda = 0.050625$. }
    \label{fig:s6f3}
\end{figure}

\begin{figure}[!ht]
    \centering
    \includegraphics[width=0.9\linewidth]{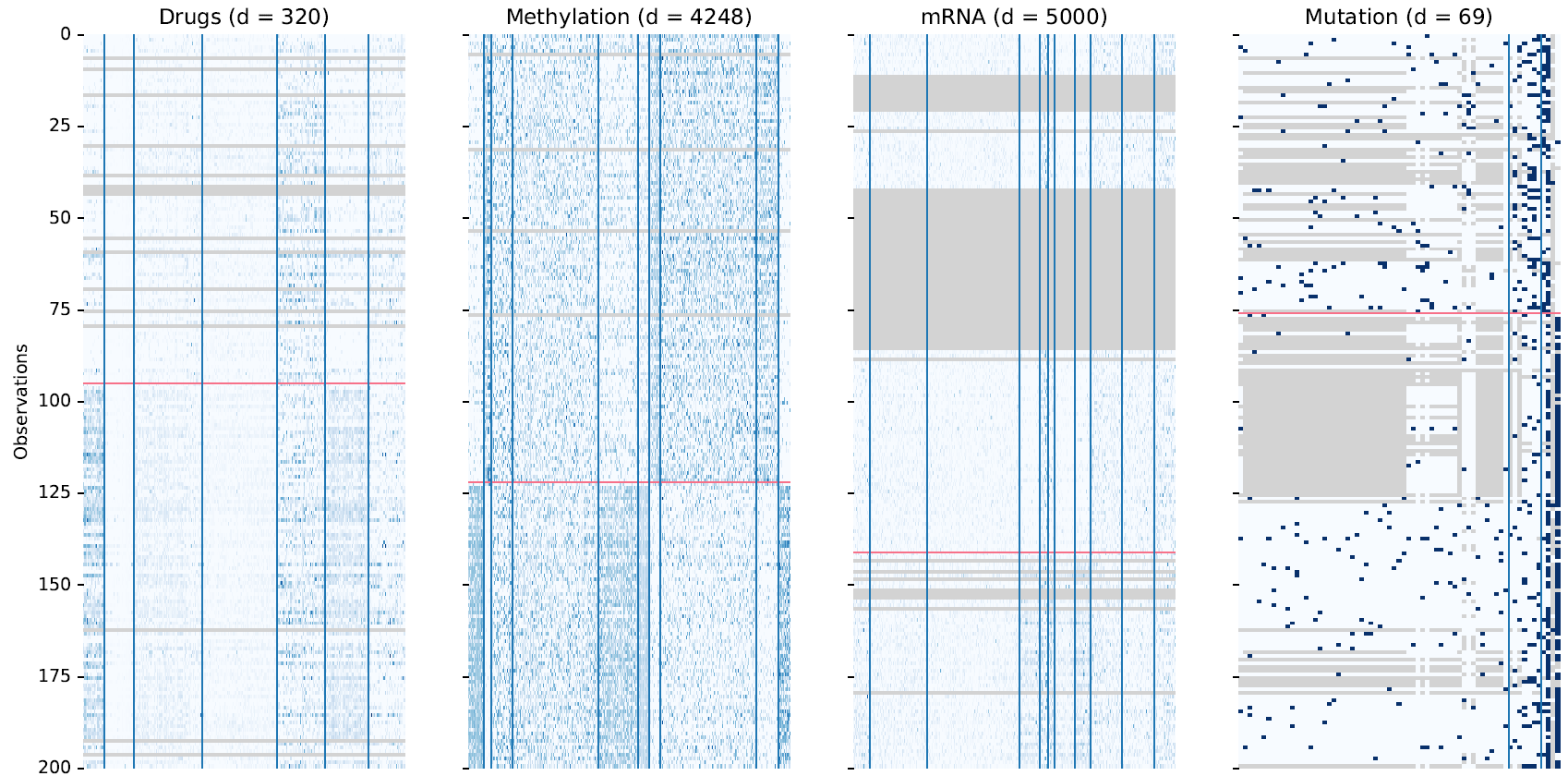}
    \caption{Overview of the clustered CLL data views. The row clusterings for each view are separated by red lines and the column clusters are separated by dark blue lines. Note that the rows are ordered independently for each view to ensure the row clusters are cohesive.}
    \label{fig:s6f4}
\end{figure}

The model selected by the ICL criterion corresponded to $\lambda = 0.050625$, which was the largest value evaluated. Notably, in 8 out of 10 simulations, the clusterings with the highest ICL took this particular penalty value. Post-penalization, only 5 clusters in the joint space were retained out of a potential 16, indicating strong relationships between the row clusterings in each view. Figure \ref{fig:s6f3} presents the marginal view of $\bm{\pi}$ and $\bm{C}$ for each pair of views. Additionally, the clusterings of each data view are provided in Figure \ref{fig:s6f4}.

The MVLBM method successfully identifies the primary clinical marker of interest in the study, IGHV mutation. Utilizing labels provided with the dataset, the row clustering returned from MVLBM achieves an ARI score of 0.79. Upon reviewing the row clustering in Figure \ref{fig:s6f4}, it is evident that the row clusterings for the Methylation and Mutation views accurately capture the IGHV mutation marker. In fact, the row clusterings are precisely common between these views, with row cluster 1 in the Methylation view corresponding exactly with row cluster 2 in the Mutation view, as illustrated in Figure \ref{fig:s6f3}.

Upon reviewing the column clusters for the Mutations view, it is evident that mutations for the Del13q14 and Gain14q32 genes, corresponding to column cluster 3, are highly predictive of mutations in IGHV status. Additionally, the presence of mutations for Trisomy12, del11q22.3, NOTCH1, and SF3B1 genes, corresponding to column cluster 2, is predictive of no mutations in IGHV status. For the Methylation column clusters, given the substantial number of features, specifying individual features can be cumbersome. However, it is noted that lower values for the methylation M-values contained in column clusters 1, 5, 6, and 10 were predictive of IGHV mutations.

Finally, it is observed that membership in the second row cluster for the Drugs view corresponds to membership in the second row cluster in the Mutations view, which is highly predictive of IGHV mutations. The first column cluster provides the greatest degree of separation for the observations in the Drugs view. This column contains four drugs: dasatinib, AT13387, PF 477736, and AZD7762, which have been shown to be predictive of IGHV mutations in \cite{argelaguet2018multi}. The ability of MVLBM to detect the primary axis of variation in the data underscores its potential for real-world applications. Moreover, MVLBM uncovers insightful information about the collected features without necessitating data transformation or the computation of composite representations.

\section{Conclusion and Future Work}\label{sec:s8}
In this work, we provided the first extension of the LBM to multi-view datasets. Our novel approach, MVLBM, is capable of simultaneously clustering the rows and columns of datasets consisting of multiple views. The method can take continuous, ordinal, categorical, and count data as an input, making it highly flexible in application. MVLBM extends the space of row clusters from single-view clusters to a joint space, where membership of a joint space cluster implies simultaneous membership of particular row clusters in each view. To determine the suitability of applying MVLBM, we introduced a hypothesis testing procedure for the null hypothesis of no relationship between the row clusters in each view. Additionally, we incorporated a log penalty scheme to encourage sparsity in the joint space of row clusters. MVLBM exhibits excellent performance in a broad range of simulated experimental analyses. Finally, we applied MVLBM to a leukaemia dataset, and demonstrated that it is capable of providing new insights for challenging genomic datasets. In future, we envisage the extension of MVLBM to incorporate other forms of data, including functional and text data, as well as exploring other schemes to induce sparsity in the joint space of clusters.

\section{Acknowledgement}
This work was funded in part by the HEA, DFHERIS and the Shared Island Fund and by the SFI grant 21/RC/10295\_P2. For the purpose of Open Access, the author has applied a CC BY public copyright licence to any Author Accepted Manuscript version arising from this submission.

\bibliographystyle{apalike}
 \bibliography{MVLBM}
\appendix
\section{Derivation of the ICL Criterion}\label{sec:a1}
We consider, for brevity, the case where $V = 2$. Using the conditional independence of $\bm{z}_1, \bm{z}_2$ and $\bm{w}_1, \bm{w}_2$ conditionally to $\bm{\Theta}$, we can formulate the integrated completed likelihood as follows. We begin by calculating $p(\bm{x}_1, \bm{x}_2, \bm{z}_1, \bm{z}_2, \bm{w}_1, \bm{w}_2 | K_1, K_2, L_1, L_2)$: 
\begin{align*}
  &\int p(\bm{x}_1, \bm{x}_2, \bm{z}_1, \bm{z}_2, \bm{w}_1, \bm{w}_2 | \bm{\alpha}_1, \bm{\alpha}_2, \bm{\pi}, \bm{\rho}_1, \bm{\rho}_2) p(\bm{\alpha}_1) p(\bm{\alpha}_2) p(\bm{\pi}) p(\bm{\rho}_1) p(\bm{\rho}_2) d\bm{\alpha}_1d\bm{\alpha}_2 d\bm{\pi} d \bm{\rho}_1 d \bm{\rho}_2 \\ 
\begin{split}
   & ={} \int p(\bm{x}_1| \bm{z}_1, \bm{z}_2, \bm{w}_1, \bm{w}_2,  \bm{\alpha}_1, \bm{\alpha}_2, \bm{\pi}, \bm{\rho}_1, \bm{\rho}_2)p(\bm{x}_2| \bm{z}_1, \bm{z}_2, \bm{w}_1, \bm{w}_2,  \bm{\alpha}_1, \bm{\alpha}_2, \bm{\pi}, \bm{\rho}_1, \bm{\rho}_2)\cdots \\
    & \quad \quad \quad \quad \quad\cdots p(\bm{z}_1, \bm{z}_2 |\bm{\alpha}_1, \bm{\alpha}_2, \bm{\pi}, \bm{\rho}_1, \bm{\rho}_2) p(\bm{w}_1, \bm{w}_2  |\bm{\alpha}_1, \bm{\alpha}_2, \bm{\pi}, \bm{\rho}_1, \bm{\rho}_2) p(\bm{\alpha}_1) p(\bm{\alpha}_2)\cdots \\
    & \quad \quad \quad \quad \quad\quad \quad\cdots p(\bm{\pi}) p(\bm{\rho}_1) p(\bm{\rho}_2) d\bm{\alpha}_1d\bm{\alpha}_2 d\bm{\pi} d \bm{\rho}_1 d \bm{\rho}_2 
\end{split}\\
\begin{split}
     & ={}  \int p(\bm{x}_1| \bm{z}_1, \bm{z}_2, \bm{w}_1, \bm{\alpha}_1,  \bm{\pi}, \bm{\rho}_1)p(\bm{x}_2| \bm{z}_1, \bm{z}_2, \bm{w}_2  , \bm{\alpha}_2, \bm{\pi},\bm{\rho}_2) p(\bm{z}_1, \bm{z}_2 |\bm{\pi})\cdots \\
    & \quad \quad \quad \quad \quad\cdots p(\bm{w}_1 |\bm{\rho}_1) p(\bm{w}_2  |\bm{\rho}_2) p(\bm{\alpha}_1) p(\bm{\alpha}_2) p(\bm{\pi}) p(\bm{\rho}_1) p(\bm{\rho}_2) d\bm{\alpha}_1d\bm{\alpha}_2 d\bm{\pi} d \bm{\rho}_1 d \bm{\rho}_2 
\end{split}\\
\begin{split}
     & ={} \int p(\bm{x}_1| \bm{z}_1, \bm{z}_2, \bm{w}_1, \bm{\alpha}_1,  \bm{\pi}, \bm{\rho}_1)p(\bm{\alpha}_1) d\bm{\alpha}_1 \int p(\bm{x}_2| \bm{z}_1, \bm{z}_2, \bm{w}_2 , \bm{\alpha}_2, \bm{\pi},\bm{\rho}_2) p(\bm{\alpha}_2) d\bm{\alpha}_2\cdots \\
    & \quad \quad \quad \quad \quad\cdots  \int p(\bm{z}_1, \bm{z}_2  |\bm{\pi}) p(\bm{\pi})  d\bm{\pi} \int p(\bm{w}_1 |\bm{\rho}_1)p(\bm{\rho}_1) d\bm{\rho}_1 \int  p(\bm{w}_2  |\bm{\rho}_2)  p(\bm{\rho}_2) d \bm{\rho}_2
\end{split}\\
& = p(\bm{x}_1 | \bm{z}_1, \bm{z}_2, \bm{w}_1) p(\bm{x}_2 | \bm{z}_1, \bm{z}_2, \bm{w}_2) p(\bm{z}_1, \bm{z}_2) p(\bm{w}_1) p(\bm{w}_2) .
\end{align*}
As such, 
\begin{equation*}
    ICL(K_1, K_2, L_1, L_2) = \log p(\bm{x}_1 | \bm{z}_1, \bm{z}_2, \bm{w}_1) + \log p(\bm{x}_2 | \bm{z}_1, \bm{z}_2, \bm{w}_2) + \log p(\bm{z}_1, \bm{z}_2) + \log p(\bm{w}_1) + \log p(\bm{w}_2) .
\end{equation*}
Following the method of \citet{keribinEstimationSelectionLatent2015}, we apply BIC-like approximations of each term of the ICL. This leads to the following approximation as $n$ and $d_1, \ldots, d_v$ tend to infinity:
\begin{align*}
 ICL ~~ \approxeq ~~ &  \log p(\bm{x}, \hat{\bm{z}}_1, \ldots,\hat{\bm{z}}_V, \hat{\bm{w}}_{1}, \ldots, \hat{\bm{w}}_{V}; \bm{\Theta}) \\  & ~~~- \frac{\sum_{v} K_{v} -1}{2} \log n \\ & ~~~~~~- \sum_{v} \frac{L_{v} - 1}{2}\log d_{v} \\ & ~~~~~~~~~ -\sum_{v}\frac{\sum_{v} K_{v} L_{v} \eta_{v}}{2} \log (n \times d_{v}),
\end{align*}
where $\eta_{v}$ is the number of parameters for each block in the $v$th view, dependent on the feature type as described in Section \ref{sec:s3s6}.

\end{document}